\documentclass[12pt]{article}
\usepackage{amssymb,amsmath,color,natbib,graphicx,
  setspace,sectsty,anysize,times,dsfont}

\marginsize{1.1in}{.9in}{.3in}{1.4in}

\newcommand{\dbl}{\setstretch{1.75}}
\newcommand{\hlf}{\setstretch{1.25}}

\newcommand{\bs}[1]{\boldsymbol{#1}}
\newcommand{\mc}[1]{\mathcal{#1}}
\newcommand{\mr}[1]{\mathrm{#1}}
\newcommand{\bm}[1]{\mathbf{#1}}
\newcommand{\ds}[1]{\mathds{#1}}

\sectionfont{\noindent\normalfont\large\bf}
\subsectionfont{\noindent\normalfont\large\textit}

\newcommand{\mT}{\mathcal{T}}

\begin{document}

\hlf

\pagestyle{empty}

\noindent {\Large \bf{Dynamic Trees for Learning and Design}}
\vskip 1cm

\noindent{\large Matthew A. Taddy, ~
Robert B. Gramacy~
and~
Nicholas G. Polson

\bigskip
\noindent {\it The University of Chicago Booth School of Business}}

\vskip 2.5cm

{\small \noindent{\sc Abstract:} Dynamic regression trees are an
  attractive option for automatic regression and classification with
  complicated response surfaces in on-line application settings.  We
  create a sequential tree model whose state changes in time with the
  accumulation of new data, and provide particle learning algorithms
  that allow for the efficient on-line posterior filtering of
  tree-states.  A major advantage of tree regression is that it allows
  for the use of very simple models within each partition. The model
  also facilitates a natural division of labor in our sequential
  particle-based inference: tree dynamics are defined through a
  few potential changes that are local to each newly arrived
  observation, while global uncertainty is captured by the ensemble of
  particles.  We consider both constant and linear mean functions at
  the tree leaves, along with multinomial leaves for classification
  problems, and propose default prior specifications that allow for
  prediction to be integrated over all model parameters conditional on
  a given tree.  Inference is illustrated in some standard
  nonparametric regression examples, as well as in the setting of
  sequential experiment design, including both active learning and
  optimization applications, and in on-line classification.  We detail
  implementation guidelines and problem specific methodology for each
  of these motivating applications. Throughout, it is demonstrated
  that our practical approach is able to provide better results
  compared to commonly used methods at a fraction of the cost.

 }

\vskip 2cm
\noindent {\footnotesize Taddy is Assistant Professor and Robert
  L. Graves Faculty Fellow, Gramacy is Assistant Professor, and Polson
  is Professor, of Econometrics and Statistics at the University of
  Chicago Booth School of Business, 5807 S Woodlawn Ave, Chicago, IL
  60637 (taddy, rbgramacy, ngp @chicagobooth.edu).  Gramacy was also a
  lecturer in the Statistical Laboratory at University of Cambridge
  during production of this article.  The research was partially
  funded by EPSRC grant EP/D065704/1 to Gramacy and by Taddy's support
  under the IBM Corporation Faculty Research Fund at Chicago
  Booth. Finally, three referees provided advice which
  greatly improved the article.  }

\newpage
\dbl

\pagestyle{plain}
\vskip 1cm
\section{Introduction}
\label{intro}

This article develops sequential regression trees, with implementation
details and examples for applications in experiment design,
optimization, classification, and on-line learning.  Our most basic
insight is the characterization of partition trees as a dynamic model
state.  This allows the model to grow with data, such that each new
observation leads to only small changes in the posterior, and has the
practical effect of reducing tree space dimension without affecting
model flexibility.  A Bayesian inferential framework is built around
particle learning algorithms for the efficient on-line filtering of
tree-states, and examples demonstrate that the proposed approach
is able to provide better results compared to commonly used methods at
a fraction of the cost.

We consider two generic regression formulations.  The first has
real-valued response (univariate here, but this is not a general
restriction) with unspecified mean function and Gaussian 
noise, and the second has a categorical response with covariate
dependent class probabilities:
\begin{equation}
a.~~~~y = f(\bm{x}) + \varepsilon,\;\;\ \varepsilon \sim \mr{N}(0,\sigma_{\bm{x}}^2)
\hskip 1.5cm
b.~~~~ \mr{p}(y=c) = p_c(\bm{x}), \;\;c=1,\ldots,C.
\label{eq:reg}
\end{equation}
The regression tree framework provides a simple yet powerful solution
to modeling these relationships in situations where very little is
known {\em a priori}.  For such models, the covariate space is
partitioned into a set of hyper-rectangles, and a simple model (e.g., linear
mean and constant error variance) is fit within each rectangle.  This
adheres to a general strategy of having a state-space model (i.e., the
partition structure) induce flexible prediction despite a restrictive
model formulation conditional on this state.  For trees, our strategy
allows posterior inference to be marginalized over the parameters of
simple models in each partition.  Given a particular tree, predictive
uncertainty is available in closed form and does not depend upon
approximate posterior sampling of model parameters.  Uncertainty
updating is then a pure filtering problem, requiring posterior
sampling for the partition tree alone.  Thereby this article
establishes a natural sequential framework for the specification and
learning of regression trees.

Although partitioning is a rough modeling tool, trees will often be
better suited to real-world applications than other ``more
sophisticated'' alternatives.  Compared to standard basis function
models for nonparametric regression, the dynamic trees
proposed herein provide some appealing properties (e.g., flexible
response surface, nonstationarity, heteroskedasiticity) at the
expense of others (e.g., smoothness, explicit correlation structure)
within a framework that is trivial to specify and allows for
conditional inference, given the global partition tree-state, to be
marginalized over all model parameters.  Prediction is very fast,
requiring only a search for the rectangle containing a new $\bm{x}$, and
individual realizations yield an easily interpretable decision tree
for regression and classification problems.  Finally, and perhaps most
importantly, the models are straightforward to fit through sequential
particle algorithms, and thus offer a major advantage in on-line
inference problems.

Consider the setting of sequential experiment design, which serves as
one of the motivators for this article. In engineering applications,
and especially for computer experiments, the typical model of
(\ref{eq:reg}.a) is built from a stationary Gaussian process (GP)
prior for $f$ with constant additive error variance \citep[see,
e.g.,][]{sant:will:notz:2003}. This GP specification is very flexible,
but inference for sample size $n$ is $O(n^3)$ due to the need to
decompose $n\times n$ matrices.  Furthermore, full Bayesian inference
is usually built around MCMC sampling, which demands $O(Bn^3)$
computation for a posterior sample of size $B$. Sequential design
requires this batch sampling to be restarted whenever new $[\bm{x},y]$
pairs are added into the data, making an already computationally
intensive procedure even more expensive by failing to take advantage
of the sequential nature of the problem.  With lower order inference
(i.e., only a search of partitions followed by simple leaf regression
model calculations) and an explicit dynamic structure, our dynamic
regression trees are less expensive and can be fit in serial.

Furthermore, it is essential that the fitted model is appropriate for
the problem of sequential design.  In this, the standard choice of a
stationary GP will again often fall short.  Whether designing for
optimization or for general learning, the goal is to search for
distinctive areas of the feature space that warrant further
exploration (e.g., due to high variance or low expected response).
The single most appropriate global specification may be very different
from the ideal local model in areas of interest (e.g., along breaks in
the response surface or near optima), such that it is crippling to
assume stationarity or constant error variance.  However,
nonstationary GP modeling schemes \citep[e.g.][]{higd:swal:kern:1999}
require even more computational effort than their stationary
counterparts.  In contrast, nonstationary function means and
nonconstant variance are fundamental characteristics of tree-based
regression.

These considerations are meant to illustrate a standard point: It may
be that some flexible functions $\{f,\sigma\}$ or $p$ are ideal for
modeling (\ref{eq:reg}.a-b) in any particular setting, but that a
simple and fast approximation -- with predictions averaged over
posterior uncertainty -- is better suited to the design scenario or
classification problem at hand.  We thus put forth dynamic
regression trees as a robust class of models with some key properties:
prior specification is scale-free and automatic; inference is
marginalized over model parameters given a global partition state; it
is possible to sequentially filter uncertainty about the partition
state; and the resulting prediction surfaces are appropriate for
modeling complicated regression functions.
 
The remainder of the paper is organized as follows.  Section
\ref{sec:trees} provides a survey of partition tree models and
\ref{sec:bayes} contains a review of Bayesian inference for trees.
Our core methodology is outlined in Section \ref{sec:seq}, which
develops a sequential characterization of uncertainty updating for
regression trees.  The partition evolution is defined in
\ref{sec:evo}, leaf model and prediction details are in
\ref{sec:leaf}, a  particle learning algorithm for posterior
simulation is provided in \ref{sec:pl}, and marginal likelihood
estimation is discussed in \ref{sec:BF}.  Section \ref{sec:examples}
then describes a set of application examples.  First, \ref{sec:reg}
illustrates linear and constant mean regression models on some simple
datasets, and compares results to alternatives from the literature.
We consider the sequential design of experiments, with
optimization search in \ref{sec:opt} and active learning in
\ref{sec:al}.  Finally, \ref{sec:class} provides two classification
examples.  Section \ref{sec:discuss} contains short closing
discussion.

\subsection{Partition Trees}
\label{sec:trees}

The use of partition trees to represent input-output relationships is
a classic nonparametric modeling technique.  A decision tree is
imposed with switching on input variables and terminal node
predictions for the relevant output.  Although other schemes are
available (e.g., based on Voronoi tessellations), the standard approach
relies on a binary recursive partitioning of input variables, as in the
classification and regression tree (CART) algorithms of
\citet*{brei:1984}.  This forces axis-aligned partitions, although
pre-processing data transformations can be used to alter the partition
space.  In general, the computational and conceptual simplicity of
rectangular partitions will favor them over alternative schemes.  As
binary recursive partition trees are fundamental to our regression
model, we shall outline some notation and details here.

Consider covariates $\bm{x}^t\!=\! \{\bm{x}_s\}_{s=1}^t$.  A
corresponding {\em tree} $\mT$ consists of a hierarchy of {\em nodes}
$\eta \in \mT$ associated with different subsets of $\bm{x}^t$.  The
subsets are determined through a series of splitting rules, and these
rules also dictate the terminal node associated with any new
$\tilde{\bm{x}}$.  Every tree has a {\em root} node, $R_{\mc{T}}$,
which includes all of $\bm{x}^t$, and every node $\eta$ is itself a
root for a sub-tree containing nodes (and associated subsets of
$\bm{x}^t$) below $\eta$ in the hierarchical structure defined by
$\mT$.  A node is positioned in this structure by its {\em depth},
$D_\eta$, defined as the number of sub-trees of $\mT$ other than $\mT$
which contain $\eta$ (e.g., $D_{R_\mc{T}} = 0$).

\begin{figure}[t!]
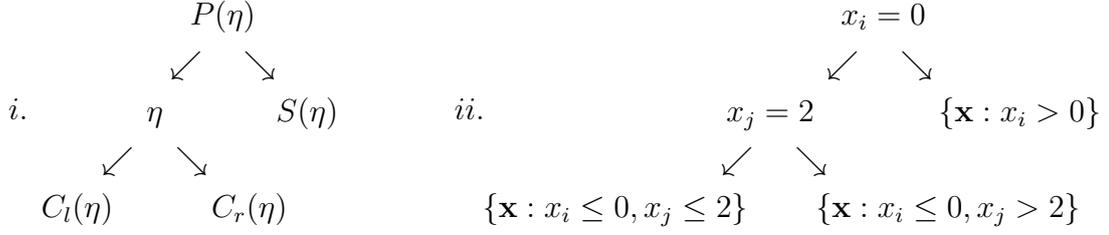

\vskip -.75cm
\hlf
\begin{equation*}
\hskip .75cm i.
\begin{array}{ccccc}
& &  \hskip -1cm P(\eta) &&\\
& & \hskip -1cm \swarrow ~~~\searrow &&\\
&  \hskip -.5cm\eta & & \hskip -.5cm S(\eta)&\\
& \hskip -.5cm \swarrow~~~\searrow & & &\\
C_{l}(\eta) & & \hskip -.3cm  C_r(\eta) & &
\end{array}
\hskip 1cm ii.\hskip -.2cm 
\begin{array}{ccccc}
& &  \hskip -2.2cm x_i = 0 &&\\
& & \hskip -2.3cm \swarrow ~~~\searrow &&\\
&  \hskip -.8cm x_j = 2 & & \hskip -2.3cm \{\bm{x}: x_i>0\}&\\
& \hskip -.7cm \swarrow~~~\searrow & & &\\
\{\bm{x}: x_i \leq  0, x_j \leq  2\}& & \hskip -.5cm \{ \bm{x}: x_i \leq 0, x_j>2 \}& &
\end{array}
\end{equation*}
\caption{ \label{tree} \small Tree heuristics.  Graph $(a)$ shows the
  parent-sibling-child network, and $(b)$ illustrates partitioning on
  two hypothetical covariates (internal nodes show split points, and
  leafs show the node set).}
\end{figure}

Figure \ref{tree} shows two diagrams of local tree structure
(partition trees tend to grow upside-down).  Graph ($i$) illustrates
the potential neighborhood of each node $\eta \in \mT$, and graph
($ii$) shows a hypothetical local tree formed through recursive binary
partitioning. Left and right {\em children}, $C_l(\eta)$ and
$C_r(\eta)$, are disjoint subsets such that $C_l(\eta) \cup C_r(\eta)
= \eta$; and the {\em parent} node, $P(\eta)$, contains both $\eta$
and its sibling node $S(\eta)$, where $\eta \cap S(\eta) = \emptyset$
and $\eta \cup S(\eta) = P(\eta)$.  Node {\em ancestors} are parents
which contain a given node, and a node's depth is equivalent to its
number of ancestors.  Any of the {\em neighbors} in ($i$) may be
nonexistent for a specific node; for example, root $R_{\mc{T}}$ has no
parent or sibling.  If a node has children it is considered {\em
  internal}, otherwise it is referred to as a {\em leaf}
node. 
The set of internal nodes for $\mc{T}$ is denoted $I_\mc{T}$, and the
set of leaves is $L_\mc{T}$.

\dbl

The tree is completed with a decision rule (i.e., a simple regression
or classification model) at each leaf. Suppose that every covariate
vector $\bm{x}_s$ is accompanied by response $y_s$, such that the
complete data is $[\bm{x},y]^t= \{\bm{x}_s,y_s\}_{s=1}^t$.  Then, with
regression models parametrized by $\theta_\eta $ for each leaf $\eta
\in L_\mT$, independence across tree partitions leads to likelihood $
\mr{p}(y^t\mid \bm{x}^t, \mT, \bs{\theta}) = \prod_{\eta\,\in\,L_\mT}
\mr{p}\left(y^\eta\mid \bm{x}^\eta, \theta_\eta\right)$, where 
$[ \bm{x}, y]^\eta = \{ \bm{x}_i, y_i : \bm{x}_i \in \eta \}$ is the data
  subset allocated to $\eta$.

\subsection{Inference for Tree Models}
\label{sec:bayes}

A novel approach to regression trees was pioneered by Chipman, George,
and McCulloch (CGM: 1998,
2002)\nocite{chip:geor:mccu:1998,chip:geor:mccu:2002}, who designed a
prior distribution, $\pi(\mT)$, over possible partition structures.
This allows for coherent inference via the posterior, $\mr{p}(\mT \mid
[\bm{x}, y]^t) \propto \mr{p}(y^t\mid\mT, \bm{x}^t) \pi(\mT)$,
including the assessment of partition uncertainty and the calculation
of predictive bands.

The CGM tree prior is generative, in that it specifies tree
probability by placing a prior on each individual partition rule.  Any
given leaf node $\eta$ may be split with depth-dependent probability
$p_{\mathrm{split}}(\mT, \eta) = \alpha(1+D_\eta)^{-\beta}$, where
$\alpha, \beta > 0$.  The coordinate (i.e. dimension of $\bm{x}$) and
location of the split, $(i,x)_\eta$, have independent prior
$p_{\mathrm{rule}}(\mT, \eta)$, which is typically a discrete uniform
distribution over all potential split points in $\bm{x}^\eta =
\bm{x}^t \cap \eta$.  Implicit in the prior is the restriction that a
partition may not be created if it would contain too few data points
to fit the leaf model (i.e., $\geq 3$ for the constant model and $\geq
d+2$ for a linear model with $d$ covariates).  Ignoring invalid
partitions, the joint prior is thus
\begin{equation}
  \pi(\mT) \propto \prod_{\eta\,\in\,\mc{I}_\mT} 
  p_{\mathrm{split}}(\mT, \eta)
  \prod_{\eta\,\in\,L_\mT}
  [1-p_{\mathrm{split}}(\mT, \eta)].
\label{eq:tprior}
\end{equation}
That is, the tree prior is the probability that internal nodes have
split and leaves have not.  

In their seminal paper, CGM develop a Metropolis-Hastings MCMC
approach for sampling from the posterior distribution of partition
trees.  This algorithm is able to explore posterior tree space by
stochastically proposing incremental modifications to $\mT$ ({\em
  grow}, {\em prune}, {\em change}, and {\em swap} moves) which are
accepted according to the appropriate Metropolis-Hastings ratio.
Although this approach provides a search of posterior trees, the
resulting Markov chain will generally have poor mixing properties.  In
particular, the proposed tree moves are very local.  An improbable
sequence of (first) {\em prunes} and (then) {\em grows} are required
to move between parts of tree space that have similar posterior
probability, but yet represent drastically different hierarchical
split rules.  Furthermore, MCMC is a batch algorithm and must be
re-run for each new observation, making it inefficient for on-line
applications.

The next section reformulates regression trees as a dynamic
 model.  Our novel characterization leads to an entirely
new class of models, and we develop a sequential particle
inference framework that is able to alleviate many of the difficulties
with MCMC for trees.

\section{Dynamic Regression Trees}
\label{sec:seq}

We now redefine partition trees as a dynamic model for predictive
uncertainty.  We introduce model state $\mc{T}_t$ which,
following from Section \ref{sec:trees}, includes the recursive
partitioning rules associated with $\bm{x}^t$, the set of covariates
observed up-to time $t$.  Section \ref{sec:evo} defines the mechanisms
of state transition $\mT_{t-1} \rightarrow \mT_{t}$ as a function of
$\bm{x}_t$, the newly observed covariates, such that $\mT_{t}$ is only
allowed to evolve from $\mT_{t-1}$ through a small set of operations
on partition structure in the neighborhood of $\bm{x}_t$ (see
Figure \ref{tree}). That is, we specify a prior distribution for the
evolution, $\mr{p}(\mT_{t} \mid \mT_{t-1}, \bm{x}^t)$.  Section
\ref{sec:leaf} then shows how to obtain tree likelihood
$\mr{p}(y^t\mid \mT_t, \bm{x}^t) $ as a product of
leaf node marginal likelihoods, and hence allows us to assign
posterior weight over the discrete set of potential trees generated
through the evolution prior.  Section \ref{sec:leaf} also details the
conditional predictive distribution and describes model particulars
for three regression leaves: constant, linear, and multinomial.  A
particle learning algorithm for on-line posterior simulation is
outlined in Section \ref{sec:pl}.  In this, each particle consists of
a tree (partitioning rules) and sufficient statistics for leaf
node predictive models, and our filtering update at time $t$ combines
small tree changes around $\bm{x}_t$ with a particle resampling
step that accounts for global uncertainty.  Finally, we describe
marginal likelihood estimation in Section
\ref{sec:BF}.

\subsection{Partition Evolution}
\label{sec:evo}

This section will specify tree dynamics through the evolution equation
$\mr{p}(\mT_{t} \mid \mT_{t-1}, \bm{x}^t)$.  Keeping possible
changes localized in the region relevant to a new observation
$\bm{x}_{t}$, the tree evolution $\mT_{t-1} \rightarrow \mT_{t}$ is
defined through one of three equally probable moves on leaf node
$\eta(\bm{x}_{t})$:
\begin{itemize}
\item[]{ \bf  stay:}
The tree hierarchy remains unchanged, and $\mT_{t}  = \mT_{t-1}$.
\item[]{\bf prune:} The tree may be {\em pruned} by removing
  $\eta(\bm{x}_{t})$ and all of the nodes below and including its
  sibling $S(\eta(\bm{x}_{t}))$.  Hence, parent $P(\eta(\bm{x}_{t}))$
  becomes a leaf node in $\mT_{t}$. If $\mT_{t-1}$ is null (contains
  only a root), then the {\em prune} move is invalid since
  $\eta(\bm{x}_{t})$ is parentless.
\item[]{\bf grow:} The {\it grow} move creates a new partition within
  the hyper-rectangle implied by split rules of the ancestors of
  $\eta(\bm{x}_t)$.  It consists of uniformly choosing split dimension
  $j$ and split point $x^{\mr{grow}}_j$ (within an interval determined
  by $\{j, \eta(\bm{x}_t), \bm{x}_t\}$ as described below), and then
  dividing $\eta(\bm{x}_{t})$ according to this rule.  The former leaf
  containing $\bm{x}_{t}$ becomes an internal node in $\mT_{t}$,
  acting as parent to two new leaf nodes (one of which contains
  $\bm{x}_{t}$).
\end{itemize}
In this way, tree changes are restricted to the neighborhood of
$\eta(\bm{x}_{t})$ illustrated in Figure \ref{tree}; we will see in
Section \ref{sec:pl} that this feature is key in limiting the cost of
posterior simulation.

Given these three possible moves, we can now define an evolution prior
as the product of two parts: a probability on each type of tree move
and a distribution over the resultant tree structure.  In the former
case, we assume that possible moves among {\it stay}, {\it prune}, and
{\it grow} are each {\it a priori} equally likely (e.g.,
$p_{\mr{move}} = 1/3$ for each when all are available), although one
may use other schemes if desired.  For the latter distribution, we
build on the inferential framework of Section \ref{sec:bayes} and
assume a CGM prior for tree structure such that $\pi(\mT_t)$
is as in (\ref{eq:tprior}) based on $p_{\mr{split}}(\mT_t, \eta) =
\alpha(1+D_\eta)^{-\beta}$.  We then have $\mr{p}(\mT_{t} \mid
\mT_{t-1}, \bm{x}^t) \propto \sum_{m \in \mc{M}(\bm{x}^t)} p_m
\pi( \mT^m)$, where $\mT^m$ is the tree that results from
applying move $m$ to $\mT_{t-1}$ and $\mc{M}(\bm{x}^t)$ is the set of
possible moves.  Hence, one may view $\mr{p}(\mT_{t} \mid
\mT_{t-1}, \bm{x}^t)$ as a covariate dependent {\it prior} for the
next tree, where the penalty on general tree shape, $\pi(\mT_{t})$,
is constant for all $t$ but the set of new potential trees you are
willing to entertain is dictated by $\bm{x}_{t}$.

It is important to note that, in contrast with staying or pruning, the
grow move actually encompasses a {\it set} of tree evolutions: we can
split on any input dimension and, for each dimension $j$, the set of
possible grow locations is the interval $\{x^\mr{grow}_j:
l^{\eta(\bm{x}_t)}_j \leq x^\mr{grow}_j \leq u^{\eta(\bm{x}_t)}_j\}$
where $l^{\eta(\bm{x}_t)}_j$ and $u^{\eta(\bm{x}_t)}_j$ are $\sup$ and
$\inf$ points such that a minimal number of observations in $
\eta(\bm{x}_{t}) \cup \{\bm{x}_{t}\}$ have $j^{\mathrm{th}}$
coordinate less than $l^{\eta(\bm{x}_t)}_j $ {\em and} strictly
greater than $u^{\eta(\bm{x}_t)}_j $ (i.e., such that each new leaf
satisfies minimum data restriction for the leaf regression model, as
described in Section \ref{sec:leaf}).  Combining these grow mechanics
with our discussion on prior specification, the implied conditional
prior for moving from $\mT_{t-1}$ to a specific {\it grown} tree
$\mT_{t}$ is the product of $p_{\mr{grow}}$ (i.e., usually $1/3$), the
penalty term $\pi(\mT_{t})$, and the probability of the specific split
location used in this grow move.  Viewed another way, the marginal
prior probability of growing in any manner is $p_\mr{grow}
\pi(\mT_{t})$ integrated over $\mT_{t}$ with respect to the measure
for possible split locations.

We place a conditional uniform prior distribution over the above set
of candidate split locations.  Hence, $j$ is uniformly distributed
over covariate dimensions with non-empty sets of possible grow points
and, given $j$, the split point is assigned a uniform prior on
$[l^{\eta(\bm{x}_t)}_j, u^{\eta(\bm{x}_t)}_j]$.  This distribution is
uniform over eligible covariate dimensions, regardless of variable
scaling within each dimension.  As an aside, since the likelihood is
unchanged for all grow moves which result in equivalent new leaf nodes
-- that is, the model is only identified up to location sets which
separate elements of $\eta(\bm{x}_{t})$ -- an alternative prior would
restrict split locations to members of $\eta(\bm{x}_{t})$ (such that
every possible grow changes the likelihood).  We prefer the former
option since it leads to smoother posterior predictive surfaces in our
particle approximations.

The formulation in this section has many things in common with the
original CGM partitioning approach.  Indeed, both involve similar {\it
  grow} and {\it prune} moves.  However, in CGM these are merely
MCMC proposals, whereas in dynamic trees they are embedded directly
into the definition of the sequential process.  Also, analogues of
{\em change} and {\em swap} are not present in our sequential
formulation.  Although these could have been incorporated, we felt
that limiting computational cost was more desirable.  In particular,
there are two aspects of our framework -- a global particle approach
to inference and a filtered posterior that changes only incrementally
with each new observation -- which allow for the convenience of a
smaller set of tree rules.

\subsection{Prediction and Leaf Regression Models}
\label{sec:leaf}

Our dynamic tree model is such that posterior inference is driven by
two main quantities: the marginal likelihood for a given tree and the
posterior predictive distribution for new data.  This section
establishes these functions for dynamic trees, and quickly details
exact forms for our three simple leaf regression models.  

First, with
each leaf $\eta \in L_{\mT_t}$ parametrized by $\theta_\eta
\stackrel{\mathrm{iid}}{\sim} \pi(\theta)$, the likelihood function is
available after marginalizing over regression model parameters as
\begin{equation}\label{eq:tlik} 
\mr{p}\left(y^t\mid \mT_t, \bm{x}^t\right) 
= \prod_{\eta\,\in\,L_{\mT_t}} \mr{p}(y^\eta\mid \bm{x}^\eta)
=  \prod_{\eta\,\in\,L_{\mT_t}} \int
\mr{p}\left(y^\eta\mid \bm{x}^\eta, \theta_\eta\right)d\pi(\theta_\eta).
\end{equation}
This is combined with the conditional prior of Section
\ref{sec:evo} to obtain posterior $\mr{p}(\mT_t \mid[\bm{x}, y]^t,\mT_{t-1})$.
Second, the predictive distribution for $y_{t+1}$ given
$\bm{x}_{t+1}$, $\mc{T}_t$, and data $[\bm{x},y]^t$, is defined
\begin{align}\label{eq:pred}
 \mr{p}(y_{t+1}\mid \bm{x}_{t+1}, \mc{T}_t, [\bm{x},y]^t)  
&=  \mr{p}(y_{t+1}\mid \bm{x}_{t+1},  [\bm{x},y]^{\eta(\bm{x}_{t+1})})
\\
& = \nonumber
  \int \mr{p}(y_{t+1} \mid \bm{x}_{t+1}, \theta )
  d\mr{P}\!\left(\theta\mid [\bm{x},y]^{\eta(\bm{x}_{t+1})}\right) 
\end{align}
where node $\eta(\bm{x}_{t+1}) \in L_{\mc{T}_t}$ is the leaf partition
containing $\bm{x}_{t+1}$ (obtained by passing through the split rules
at each level of $\mT_t$) and $d\mr{P}$ is the posterior distribution
over leaf parameters given the data in $\eta(\bm{x}_{t+1})$.  That is,
the predictive distribution for covariate vector $\bm{x}$ is just the
regression function at the leaf containing $\bm{x}$, integrated over
the conditional posterior for model parameters.  In a somewhat subtle
point, we note that a modeling choice has been made in our statement
of (\ref{eq:pred}): prediction at $\bm{x}_{t+1}$ depends only on
$\mT_{t}$, rather than averaging over potential $\mT_{t+1}$ as would
be necessary for a conventional state-space model.  Our predictive
function in (\ref{eq:pred}) leads to substantially more efficient
inference (the alternative is impractical in high dimension), and it
seems to us that the distinction should not produce dramatically
different results.

We concentrate on three basic options for leaf regression: the
canonical constant and linear mean leaf models, and multinomial leaves
for categorical data.  From (\ref{eq:tlik}) and (\ref{eq:pred}), we
see that conditioning on a given tree reduces our necessary posterior
functionals to the product of independent leaf likelihoods and a
single leaf regression function, respectively.  Ease of implementation
and general efficiency of our models depends upon an ability to
evaluate analytically the integrals in (\ref{eq:tlik}) and
(\ref{eq:pred}), such that prediction and likelihood are always
marginalized over unknown regression model parameters (given a default
scale-free prior specification).  Fortunately, such quantities are
easily available for each leaf, and the following three subsections
quickly outline modeling specifics for our three types of regression
tree.

Before moving to leaf details, we note that the complexity of leaves
is limited only by computational budget and data dimensions.  For
example, the treed Gaussian process (TGP) approach of
\citet{gra:lee:2008} fits a GP regression model at each leaf node.
The tree imposes an independence structure on data covariance,
providing an inexpensive nonstationary GP model.  However, unlike the
models proposed herein, GP leaves do not allow inference to be
integrated over regression model parameters.  This significantly
complicates posterior inference, leading to either reversible-jump
methods for MCMC or much higher-dimensional particles for sequential
inference (which can kill algorithm efficiency).  Thus, although
dynamic trees can be paired with more complex models, the
robust nature of less expensive trees will lead to them being the
preferred choice in many industrial applications.  Indeed, our results
in Section \ref{sec:examples} indicate that extra modeling does not
necessarily lead to better performance.

\subsubsection{Constant Mean Leaves}
\label{sec:cmt}

Consider a tree $\mc{T}_t$ partitioning data $[\bm{x},y]^t$ at time
$t$, and let $|\eta|<t$ be the number of observations in
leaf node $\eta$.  The constant mean model assumes that $y^\eta =
\{ y_i : \bm{x}_i \in \eta\}$ are distributed as
\begin{equation}\label{cmod}
y^\eta_1,\dots, y^\eta_{|\eta|}
\stackrel{\mathrm{iid}}{\sim} \mr{N}(\mu_\eta, \sigma_\eta^2).
\end{equation}
Under this model, leaf sufficient statistics are $ \bar{y}_\eta
=\sum_{i=1}^{|\eta|} y^\eta_i/|\eta|$ and $s_\eta^2=
\sum_{i=1}^{|\eta|} (y^{\eta}_i - \bar{y}_\eta)^2 =
\sum_{i=1}^{|\eta|} (y^{\eta}_i)^2 - |\eta|\bar{y}_\eta^2$, where our
prior forces the minimum data condition $|\eta|>2$.  Note that these
statistics are easy to update when the leaf sets change.

Under the motivation of an {\it automatic} regression framework, we
assume independent scale-invariant priors for each leaf model, such
that $\pi(\mu_\eta, \sigma_\eta^2) \propto 1/\sigma_\eta^2$.  The leaf
likelihood is then
\begin{align}
  \mr{p}(y^\eta\mid \bm{x}^\eta) = \mr{p}(y^\eta) &= \int
  \mr{N}(y^\eta\mid\mu_\eta, \sigma_\eta^2)
  \frac{1}{\sigma_\eta^2}\; d\mu_\eta d\sigma_\eta \nonumber \\
  &= \frac{1}{(2\pi)^{\frac{|\eta|-1}{2}}}\frac{1}{\sqrt{|\eta|}}
  \left(\frac{s_\eta^2}{2}\right)^{-\frac{|\eta|-1}{2}}
  \Gamma\left(\frac{|\eta|-1}{2}\right).
\label{eq:cmrg}
\end{align}
For
covariate vector $\bm{x}$ allocated to leaf node $\eta$, the posterior
predictive distribution is
\begin{equation}
  \mr{p}(y\mid \bm{x}, \eta, [\bm{x},y]^\eta) = \mr{p}(y\mid \eta, y^\eta)  
= \mr{St}\left(y;~\bar{y}_{\eta},~
  \frac{\left( 1 + \frac{1}{|\eta|} \right)}{ 
    |\eta| - 1 }s_{\eta}^2,~ |\eta|-1 \right),
\label{eq:cpred}
\end{equation}
where $\mr{St}$ is a Student-$t$ distribution with $|\eta|-1$ degrees of
freedom.

\subsubsection{Linear Mean Leaves}
\label{sec:lmt}

Consider extending the above model to a linear leaf regression
function.  Leaf responses $y^\eta$ are accompanied by design matrix
$\bm{X}_\eta = [\bm{x}^\eta_{1},\dots,\bm{x}^\eta_{|\eta|}]^\prime$,
where $\{\bm{x}_i^\eta\}_{i=1}^{|\eta|} = \bm{x}^t \cap \eta$, and
\begin{equation}\label{lmod}
y^\eta \sim \mr{N}(\mu_\eta1_{|\eta|} +\bm{X}_{\eta}\beta_\eta,
\sigma_\eta^2\bm{I}_{|\eta|})
\end{equation}
for intercept parameter $\mu_\eta$, $d$-dimensional slope parameter
$\beta_\eta$, and error variance $\sigma_\eta^2$.  Sufficient
statistics for leaf node $\eta$ are then $\bar{y}_\eta$ and
$s^2_\eta$, as in Section \ref{sec:cmt}, plus the covariate mean
vector $\bar{\bm{x}}_\eta = \sum_{i=1}^{|\eta|} \bm{x}^\eta_i/|\eta|$,
shifted design matrix $\hat{\bm{X}}_\eta =
[\bm{x}^\eta_{1}-\bar{\bm{x}}_\eta,\dots,\bm{x}^\eta_{|\eta|}-\bar{\bm{x}}_\eta]^\prime$,
Gram matrix $\mc{G}_\eta = \hat{\bm{X}}_\eta^\prime\hat{\bm{X}}_\eta$
and 
slope vector $\hat{\beta}_\eta = \mc{G}_\eta^{-1}
(\hat{\bm{X}}_\eta)^\prime({y}^\eta-\bar{y}_\eta)$ for the
  shifted design matrix, and regression sum
of squares $\mc{R}_\eta = \hat{\beta}_\eta^\prime \mc{G}_\eta
\hat{\beta}_\eta$.  We now restrict $|\eta|\geq d+2$.  As before,
these statistics are easily updated; in particular, matrices can be
adapted through partitioned inverse equations.

We again assume the scale invariant prior $\pi(\mu_\eta, \beta_\eta,
\sigma^2_\eta) \propto 1/\sigma_\eta^2$ and, as above, it is
straightforward to calculate the essential predictive probability and
marginal likelihood functions.  The marginal likelihood for leaf node
$\eta$ is then
\begin{equation}\label{eq:lmrg}
\mr{p}(y^\eta | \bm{x}^{\eta}) 
= \frac{1}{(2\pi)^{\frac{|\eta|-d-1}{2}}}
\left(\frac{|\mc{G}_\eta^{-1}|}{|\eta|}\right)^{\frac{1}{2}}
\left(\frac{s_\eta^2-\mc{R}_\eta}{2}\right)^{-\frac{|\eta|-d-1}{2}} 
\Gamma\left(\frac{|\eta|-d-1}{2}\right).
\end{equation}
For $\bm{x}$ allocated to leaf node
$\eta$ and with $\hat{\bm{x}} = \bm{x}-\bar{\bm{x}}_\eta$, the posterior
predictive distribution is 
\begin{equation}\label{eq:lpred}
  \mr{p}(y \mid \bm{x}, \eta, [\bm{x},y]^\eta )
  = \mr{St}\!\left(y;~
    \bar{y}_\eta +\hat{\bm{x}}^\prime \hat{\beta}_\eta,~
    (1\!+\!|\eta|^{-1}\!+ \hat{\bm{x}}^\prime 
    \mc{G}_\eta^{-1}\hat{\bm{x}})\! 
    \left[\frac{s_\eta^2 - \mc{R}_\eta}{|\eta|\!-\!d\!-\!1}\right]\!,~ |\eta|\!-\!d\!-\!1 \right)\!.
\end{equation}

\subsubsection{Multinomial Leaves}
\label{sec:mnl}

For categorical data, as in (\ref{eq:reg}.b), each leaf response
$y^\eta_s$ is equal to one of $C$ different factors.  The set
of responses, $y^\eta$, can be summarized through a count vector
$\bm{z}_\eta = [z^\eta_1,\ldots,z^\eta_C]^\prime$, such that $z^\eta_c =
\sum_{s=1}^{|\eta|} \ds{1}({y^\eta_s=c})$.  Summary
counts for each leaf node are then modeled as
\begin{equation}\label{clmod}
\bm{z}_\eta \sim \mr{MN}( \bs{p}_{\eta}, |\eta|)
\end{equation}
where $\mr{MN}( \bs{p}, n)$ is a multinomial with expected count
$p_c/n$ for each category. 

We assume a default Dirichlet $\mr{Dir}( 1_{C}/C)$ prior for each leaf
probability vector, such that posterior information about
$\bs{p}_{\eta}$ is summarized by $\bs{\hat{p}}_\eta =
(\bm{z}_{\eta}+1/C)/(|\eta|+1)$, which is trivial to update.
 The marginal likelihood for leaf node $\eta$ is then
\begin{equation}\label{eq:clmrg}
  \mr{p}(y^\eta\mid \bm{x}^\eta) = \mr{p}(\bm{z}_\eta) 
= \frac{1}{|\eta|!}\prod_{c=1}^C\frac{\Gamma(z^\eta_c +
  1/C)}{\Gamma(1/C)}.
\end{equation}
Covariates $\bm{x}$ allocated to leaf node $\eta$ lead to
predictive response probabilities
\begin{equation}\label{eq:clpred}
  \mr{p}(y=c \mid \bm{x}, \eta, [\bm{x},y]^\eta) = \mr{p}(y=c\mid
  \bm{z}_\eta)  = \hat{p}_c^\eta,  ~\text{for}~c=1,\ldots,C.
\end{equation}

\subsection{Particle Learning for Posterior Simulation}
\label{sec:pl}

Posterior inference about dynamic regression trees is obtained through
sequential filtering of a representative sample.  The posterior
distribution over trees at time $t-1$ is characterized by $N$ equally
weighted particles, each of which includes a tree $\mT_{t-1}^{(i)}$
encoding recursive partitioning rules and $S^{(i)}_{t-1}$, the
associated set of all sufficient statistics for each leaf regression
model.  Upon the arrival of new data, particles are first resampled
and then propagated to move the sample from $\mr{p}\left( [\mT,
  S]_{t-1}\mid [\bm{x},y]^{t-1}\right)$ to $\mr{p}\left( [\mT, S]_{t}
  \mid [\bm{x},y]^{t}\right)$.

In particular, our approach is a version of the particle learning (PL)
sequential Monte Carlo algorithm introduced by
\citet{CarvJohaLopePols2009} in the context of mixtures of dynamic
linear models. The methods in this section are specific to dynamic
regression trees and we defer further detail to the original PL paper
\citep[in addition,][focuses on PL for general mixtures and thus
describes a partition model with parallels to
trees]{CarvLopePolsTadd2009}.  In the context of our dynamic trees, PL
recursive update equations are
\begin{eqnarray}\label{eq:pl}
  \mr{p}\left([\mT, S]_t \mid [\bm{x},y]^t\right)  &=& 
  \int \mr{p}\left([\mT, S]_t\mid [\mT, S]_{t-1}, [\bm{x},y]_t\right)
  d\mr{P}\left([\mT, S]_{t-1} \mid [\bm{x},y]^t\right) \\
  &\hskip -5.5cm \propto& \hskip -3cm \int \mr{p}\left( [\mT, S]_t \mid [\mT, S]_{t-1}, [\bm{x},y]_t\right) 
  \mr{p}\left([\bm{x},y]_t \mid [\mT, S]_{t-1}\right) 
  d\mr{P}\left( [\mT, S]_{t-1} \mid [\bm{x},y]^{t-1}\right),\notag
\end{eqnarray}
where the second line is due to a simple application of Bayes rule.
Hence, the particle set $\{[\mT,S]_{t-1}^{(i)}\}_{i=1}^N \sim
\mr{p}([\mT,S]_{t-1} \mid [\bm{x},y]^{t-1})$ can be updated by first
{\it resampling} particles proportional to $\mr{p}( [\bm{x},y]_t \mid
[\mT,S]_{t-1})$, and then {\it propagating}  from
$\mr{p}([\mT,S]_t \mid [\mT,S]_{t-1}, [\bm{x},y]_t)$. 

Although wider questions of PL efficiency are addressed in
\citet{CarvJohaLopePols2009}, we should make some quick notes on the
tree-specific algorithm.  Crucially, resampling first to condition on
$[\bm{x},y]^{t}$ introduces information from $[\bm{x},y]_t$ that is
not conditional on $\mT_t$, thus reducing direct dependence of
particles on the high dimensional tree-history $\mT^t$ (this same idea
is found in the best particle filtering algorithms, such as
\citet{KongLiuWong1994} and \citet{PittShep1999}, even if it remains
standard to track and weight all of $\mT^t$).  Second, any analytical
integration over model parameters will greatly improve Monte Carlo
sampling performance through more efficient Rao-Blackwellized
inference.  In our case, simple leaf models allow direct evaluation of
$ \mr{p}(\mT_t | [\bm{x},y]^t) = \int \mr{p}( \mT_t,
\{\theta_\eta:\eta \in L_{\mT_t} \}| [\bm{x},y]^t)
d\mr{P}(\{\theta_\eta:\eta \in L_{\mT_t} \} | [\bm{x},y]^t)$; without
such marginalization, particles would need to be augmented with all
leaf parameters.  Finally, since only the neighborhood of
$\eta(\bm{x}_{t})$ changes under any of these moves, it is possible
to ignore all other data and tree structure when calculating relative
probabilities for tree changes.  This, combined with grow moves that
are only identified up to a discrete set of locations, restricts
particle updates to a low dimensional and easily sampled space.

In full detail, suppose that $\{\mT_{t-1}^{(i)}\}_{i=1}^N \sim
\mr{p}(\mT_{t-1} \mid [\bm{x},y]^{t-1})$ is a particle approximation
to the posterior at time $t-1$.  The dimension of each
$[\mT,S]_t^{(i)}$ depends upon both the tree structure and the
regression model, with each particle containing $| L_t^{(i)} |$ sets
$S_{t\eta}^{(i)}$ of the leaf sufficient statistics described in
Section \ref{sec:leaf}.  All trees/particles begin empty.  Due to
minimum data restrictions for each partition, {\it stay} is the only
plausible tree move until enough data has accumulated to split into
two partitions.  Hence, updating begins at $t = 6$ for the constant
model and at $t = 2(d+2)$ under linear leaves, with each
$\mT_{t}^{(i)}$ consisting of a single leaf/root node.

Upon observing new covariates $\bm{x}_t$ and response $y_t$, update as follows.
\begin{itemize}
\item[] {\bf Resample:} Draw particle indices
  $\{\zeta(i)\}_{i=1}^N$ with predictive probability weight
\vskip -.75cm \[
\mr{p}(\zeta(i)=i) \propto \mr{p}(y_{t}\mid \bm{x}_{t}, [\mc{T},S]_{t-1}^{(i)})  
= \mr{p}(y_{t} \mid \bm{x}_{t}, S_{t\eta(\bm{x}_{t})}^{(i)} ),
\]\vskip -.25cm 
as in (\ref{eq:cpred}), (\ref{eq:lpred}), or (\ref{eq:clpred}).  There
are various low-variance options for resampling
\citep[e.g.,][]{cappe:douc:moulines:2005}, and our implementation
makes use of residual-resampling.

Set $\mT_{t-1}^{(i)} = \mT_{t-1}^{\zeta(i)}$ for each $i$ to form a
new particle set.

\item[] {\bf Propagate:} We need to update each tree particle with a
  sample from the discrete distribution $\mr{p}([\mT,S]_t \mid
  [\mT,S]_{t-1}, [\bm{x},y]_t) \propto \mr{p}(\mT_{t} \mid \mT_{t-1},
  \bm{x}^t)\mr{p}(y^t\mid \mT_t, \bm{x}^t)$.  First, propose changes
  for $\mT_{t-1}^{(i)} \rightarrow \mT_{t}^{(i)}$ via {\it stay}, {\it
    prune}, or a randomly sampled {\it grow} move on
  $\eta_{t-1}^{(i)}(\bm{x}_t)$.  For each particle's {\it grow}, we
  propose from the uniform grow-prior of Section \ref{sec:evo} by
  drawing $j$ from eligible dimensions (i.e. those with non-empty sets
  of split locations) and then sampling the split point from a uniform
  on $[l_j^{\eta(\bm{x}_t)},u_j^{\eta(\bm{x}_t)}]$.

  The three candidate trees are now $\mT_{t}\in \{\mT^{\mr{stay}},
  \mT^{\mr{prune}}, \mT^{\mr{grow}}\}$.

Since candidate trees are equivalent above $P(\eta(\bm{x}_{t}))$, the
parent node for $\bm{x}_{t}$ on tree $\mT_{t-1}$, we calculate
posterior probabilities only for the subtrees rooted at this node.
With candidate subtrees denoted $\mT^{\mr{move}}_t$ and containing
data $[\bm{x},y]^t$ (including $\bm{x}_{t}$ and $y_t$), the new
$\mT_{t}$ is sampled with probabilities proportional to
$\pi(\mT_{t}^\mr{move}) \mr{p}(y^t \mid \bm{x}^{t}, \mT^\mr{move}_{t})$.  Here,
the prior penalty is (\ref{eq:tprior}) and the likelihood is
(\ref{eq:tlik}) with leaf marginals from (\ref{eq:cmrg}),
(\ref{eq:lmrg}), or (\ref{eq:clmrg}).

Finish with deterministic sufficient statistic updates $S^{(i)}_{t-1} \rightarrow
S^{(i)}_{t}$.
\end{itemize} 
These two simple steps yield an updated particle approximation
$\{\mT_{t}^{(i)}\}_{i=1}^N \sim \mr{p}(\mT_{t}\mid [\bm{x},y]^{t})$.

In an appealing division of labor, resampling incorporates global
changes to the tree posterior, while propagation provides local
modifications.  As with all particle simulation methods, some Monte
Carlo error will accumulate and, in practice, one must be careful to
assess its effect.  However, as mentioned above, our strategy makes
major gains by integrating over model parameters to obtain particles
which consist of only split rules and sufficient statistics.  Given
this efficient low-dimensional particle definition, our
resampling-first procedure will sequentially discard all models except
those which are predicting well, and this tempers posterior search to
a small set of plausible trees with good predictive properties. We
will see in Section \ref{sec:examples} that PL for trees has some
significant advantages over the traditional MCMC approach.

\subsection{Marginal Likelihood Estimation}
\label{sec:BF}

An attractive feature of our dynamic trees is that, due to the use of
simple leaf regression models, reliable posterior marginal likelihood
estimates are available through the sequential factorization
$\mr{p}(y^T \mid \bm{x}^T) = \prod_{t=1}^T \mr{p}(y_t \mid \bm{x}_t,
[\bm{x},y]^{t-1})$.  However, use of improper priors in constant and
linear trees implies that the marginal likelihood is not properly
defined.  As described in \citet{Atki1978}, this can be overcome if
some of the data is set aside, before any model comparison, and used
to form proper priors for leaf model parameters.  Such ``training
samples'' are already enforced within our dynamic tree evolution
through the minimum partition-size requirements for each model.  As
such, a valid approximate marginal likelihood is available by
conditioning on the first $t_0$ observations, where $t_0$ is large
enough to provide a proper posterior predictive distribution for the
unpartitioned ``root'' tree.  Hence, we write $\mr{p}(y^T \mid
\bm{x}^T) \approx \prod_{t=t_0+1}^T \mr{p}(y_t \mid \bm{x}_t,
[\bm{x},y]^{t-1})$ where the factors are approximated via PL as
\begin{equation}\label{mlhd}
\ds{E}\left[ \mr{p}(y_t
  \mid \bm{x}_t, [\bm{x},y]^{t-1}, \mT_{t-1}) 
\mr{p}( \mT_{t-1}\mid [\bm{x},y]^{t-1}) \right]
\approx \frac{1}{N} \sum_{i=1}^N \mr{p}(y_t \mid \bm{x}_t, [\mT,S]_{t-1}^{(i)}).
\end{equation}
This is just the normalizing constant for PL resampling probabilities
(refer to Section \ref{sec:pl}).

These marginal likelihood calculations make it possible to compare
leaf model specifications through the marginal likelihood ratio or
Bayes factor (BF).  For example, in a comparison of linear leaves
against a constant leaf model, $\mr{p}_{\mr{lin}}(y^T \mid
\bm{x}^T)/\mr{p}_{\mr{const}}(y^T \mid \bm{x}^T)$ measures the relative
evidence in favor of linear leaves \citep[see][for information on
assessing BFs; also, note that the posterior probability of linear
leaves is $\mr{p}_{\mr{lin}}(y^T \mid \bm{x}^T) /(\mr{p}_\mr{lin}(y^T \mid
\bm{x}^T) + \mr{p}_\mr{const}(y^T \mid \bm{x}^T) )$ given even prior
probability for each model]{KassRaft1995}.  Linear and constant leaves
are just extreme options for the general model choice of which
covariates should be leaf regressors, and the BF approach is
generally applicable in such problems.

This approach to inference {\it is} data-order dependent, due to the
state-space factorization assumption $\mr{p}(y_t \mid \bm{x}^T,
y^{t-1}) = \mr{p}(y_t \mid \bm{x}^t, y^{t-1})$.  In addition, BFs are
only available conditional on the initial $y^{t_0}$ training sample,
and clearly for marginal likelihood estimation both training sample
and data-order must be the same across models.  However, we will see
in Section \ref{sec:reg} that these BFs lead to consistent model
choices in the face of strong evidence, and that average BFs over
repeated reorderings and training samples serve an effective model
selection criterion.

\section{Application and Illustration}
\label{sec:examples}

We now present a series of examples designed to illustrate our
approach to on-line regression.  Section \ref{sec:reg} considers two
simple 1-d regression problems, illustrating both constant and linear
mean leaf models over multiple PL runs, before comparing dynamic tree
predictive performance against competing estimators in a 5-d
application.  The next two sections focus on the sequential design of
experiments, with optimization problems in Section \ref{sec:opt} and
active learning in Section \ref{sec:al}.  Finally, Section
\ref{sec:class} describes the application of dynamic multinomial leaf
trees to a 15-d classification problem.  Throughout, our dynamic trees
were fit using the {\tt dynaTree} package \citep{dynaTree} for {\sf R}
under the default parametrization.
In particular, the tree prior of Section \ref{sec:bayes} is specified
with $\alpha=0.95$ and $\beta =2$; inference is generally robust to
reasonable changes to this parametrization (e.g., the four settings
described in Figure 3 of CGM (2002) lead to no qualitative
differences).

\subsection{Regression Model Comparison}
\label{sec:reg}

We begin by considering two simple 1-d applications.  The first ``parabola
data'' set has 100 observations from $y(x) \sim \mr{N}(x + x^2,
1/5^2)$, where $x$ was drawn uniformly from $[-3,3]$.  The second
``motorcycle data'' set,
available in the {\tt MASS} library 
for {\sf R},
consists of observed acceleration on motorcycle riders' helmets at 133
different time points after simulated impact.

\begin{figure}[p!]
\vskip -.5cm
\hskip -.5cm\includegraphics[width=6.7in]{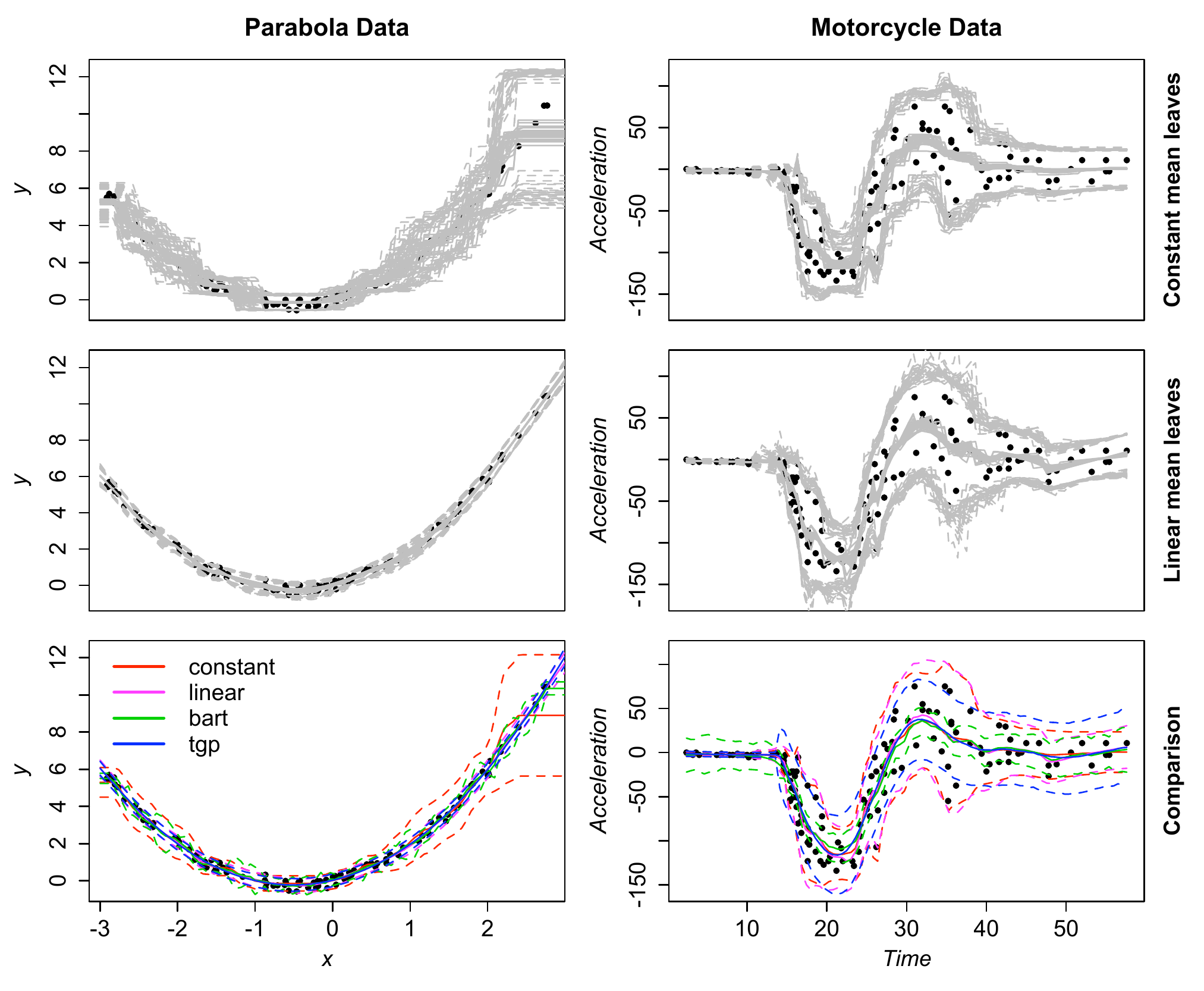}
\vskip -.5cm
\caption{\label{fig:reg} Top two rows show, for constant and linear
  leaves, posterior mean and 90\% interval for each of 30 runs with
  1000 particles to randomly ordered data. Bottom row shows mean and
  90\% interval for the average across dynamic tree runs, as well as
  for BART and TGP.}
\vskip .25cm
\hskip -.45cm\includegraphics[width=6.5in]{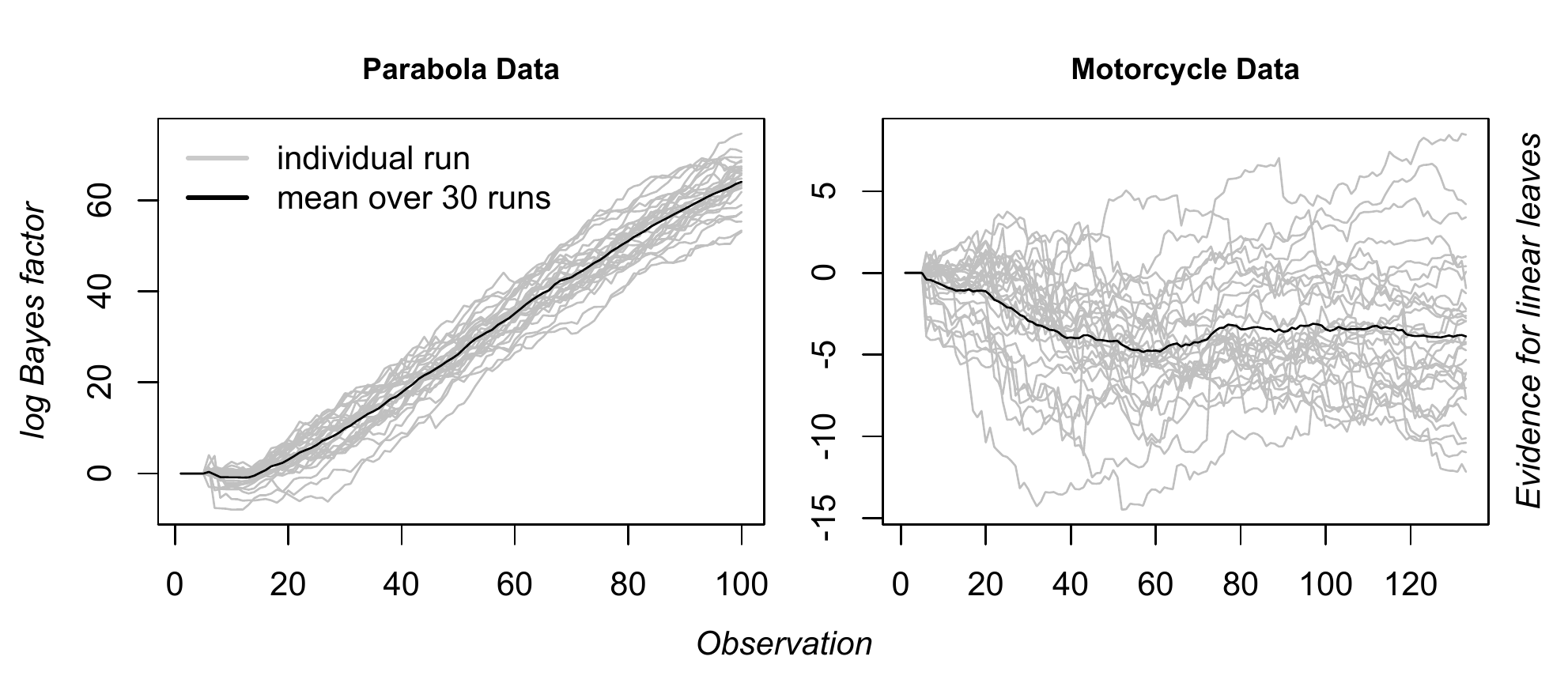}
\vskip -.5cm
\caption{\label{fig:bf} Filtered log
  Bayes factors for linear against constant leaf tree models, calculated for 30
  random data orderings (the mean is shown in black) with 1000 particles. }
\end{figure}

The top two rows of Figure \ref{fig:reg} show 30 repeated filtered
posterior fits for random re-orderings of the data, obtained through
the PL algorithm of Section \ref{sec:pl} with 1000 particles.  The
first row corresponds to constant leaf models while the second row
corresponds to linear leaves, and each plot shows posterior mean and
90\% predictive intervals.  Although there is clear variation from one
PL run to the next, this is not considered excessive given the random
data orderings and small number of particles.  Linear leaf models
appear to be far better than constant models at adapting to the
parabola data.  With the parabola data, a constant leaf model leads to
higher average tree height (9.4, vs 5.4 for linear leaves) while the
opposite is true for the motorcycle data (average constant leaf height
is 10.5, vs 12.8 for linear leaves).

A more rigorous assessment of the relative evidence in favor of each
leaf model is possible through estimated Bayes factors, as described
in Section \ref{sec:BF}.  Figure \ref{fig:bf} presents filtered log
BFs comparing linear to constant leaves, calculated for the 30 random
data orderings used in Figure \ref{fig:reg} and conditional on the
first $t_0 = 5$ observations.  As discussed in \ref{sec:BF}, due to
both random ordering and different $y^{t_0}$ training samples, the BF
estimates are not directly comparable across runs. However, for the
parabola data, in every case the linear model is clearly preferable.
For the motorcycle data, although there is no consistent evidence in
favor of either model, the majority of runs produce log BF values
below zero and the mean across runs (which eliminates order dependence
as the number of runs increases) shows fairly strong evidence against the
linear model.  This agrees with the visually similar regression fits
drawn in Figure \ref{fig:reg}, which were obtained after fewer average
partitions under constant leaves than with the linear leaves.

Finally, the bottom row of Figure \ref{fig:reg} compares the means of
the 30 dynamic tree fits to two similar modern nonparametric
regression techniques: treed Gaussian processes (TGP) and Bayesian
additive regression trees \citep[BART;][]{ChipGeorMcCu2010}.  Each new
model is designed as an extension to constant or linear regression
trees, and makes steps to partially alleviate the mixing problems of
CGM's original MCMC inference.  As mentioned previously, TGP takes
advantage of a more flexible leaf regression model to allow for
broader covariate partitions.  In a different approach, BART proposes
a mixture of relatively short trees, and the authors show that
combinations of simple individual partitioning schemes can lead to a
complicated predictive response surface.  The TGP model is given 10
restarts of a 10,000 iteration MCMC run, while BART results are based
on a single 1,100 iteration chain.  For the parabola data, all of the
models find practically identical fits, except for the poorly
performing constant leaf dynamic tree.  However, for the motorcycle
data, we see that each model leads to mean functions that are very
similar, but that posterior predictive 90\% intervals for BART and TGP
appear to variously over or under estimate data uncertainty around the
regression mean.  In particular, BART's global variance term is
misspecified for this heteroskedastic data.

To benchmark our methods on a more realistic high-dimensional example,
we also consider out-of-sample prediction for a Friedman test function
\citep[originally designed to illustrate multivariate adaptive
regression splines (MARS) in][]{fried:1991}.  The response is $10
\sin(\pi x_1 x_2) + 20(x_3 - 0.5)^2 + 10x_4 + 5 x_5$ with
$\mr{N}(0,1)$ additive error.  Our comparators, and their {\sf R}
implementations, are: treed constant (TC) and linear (TL) models, and
GP models from {\tt tgp} \citep{tgp}; MARS from {\tt mda} \citep{mda};
Random Forests (RF) from {\tt randomForest} \citep{rf}; BART from {\tt
  BayesTree} \citep{bart}; and neural networks with 2-3 hidden layers
from the {\tt nnet} library.

\begin{figure}[h!]
\centering
\begin{minipage}{8cm}
\includegraphics[scale=0.75,trim=0 30 0 30]{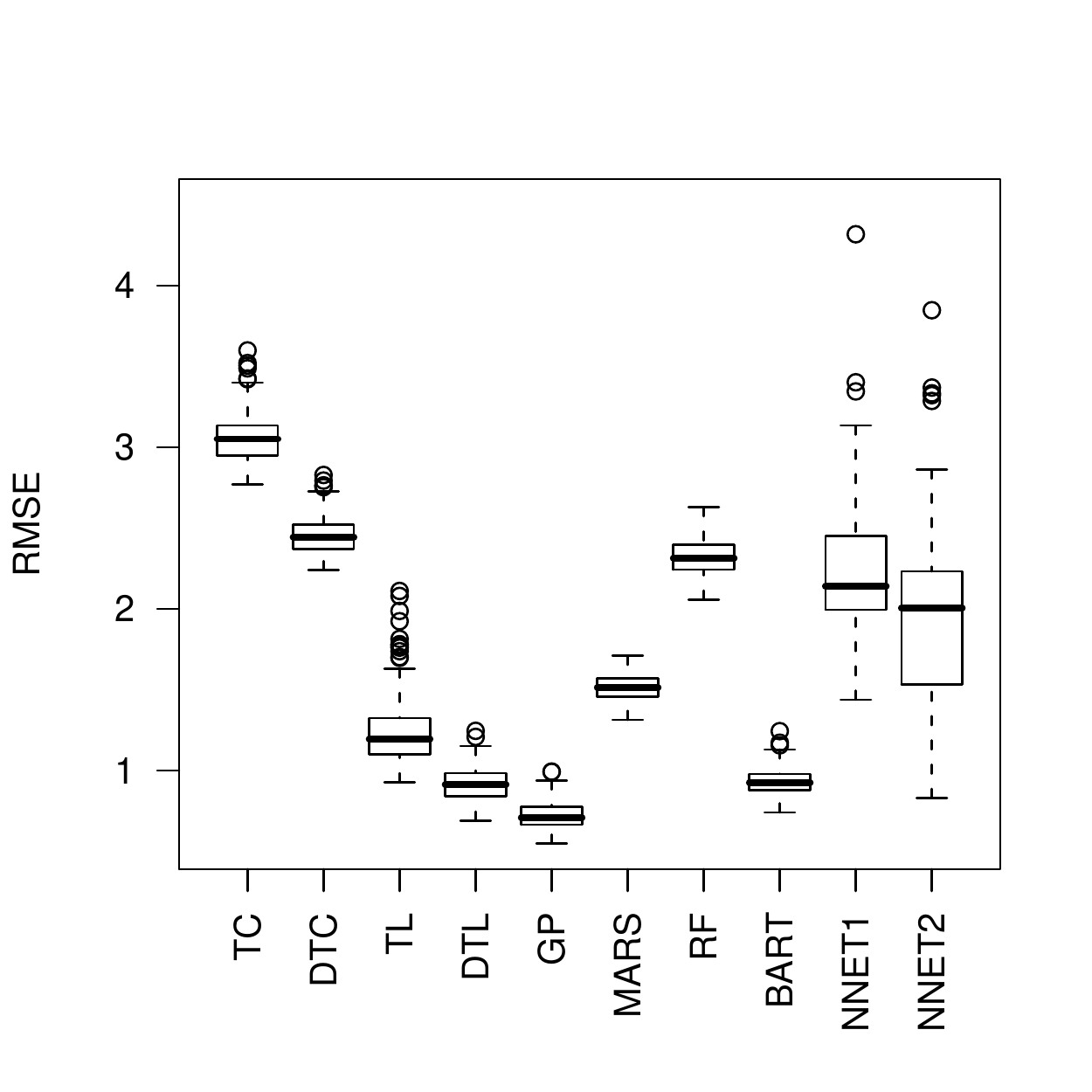}\\
\end{minipage}
\hfill
\begin{minipage}{5cm}
\begin{tabular}{l|rr}
& \multicolumn{2}{c}{RMSE} \\
method & mean & (sd) \\
\hline
GP    & 0.723 & (0.053) \\
DTL  & 0.917 & (0.104) \\
BART  & 0.935 & (0.063) \\
TL  &  1.244 & (0.217) \\
MARS & 1.513 & (0.072) \\
NNET2 & 1.913 & (0.429) \\
NNET1 & 2.246 & (0.424) \\
RF & 2.320 & (0.093) \\
DTC & 2.459 & (0.074) \\
TC & 3.067 & (0.102) 
\end{tabular}
\end{minipage}
\caption{Predictive RMSE on the Friedman data; sample of 100 for each
  estimator.}
\label{f:fried}
\end{figure}

For this experiment, 100 random training and prediction sets, of
respective size 200 and 1000, were drawn with inputs uniform on
$[0,1]^5$. Following \citet{chip:geor:mccu:2002}, we measure
performance through predictive root mean-square error (RMSE) to the
true mean, and results are shown in Figure \ref{f:fried}.  Dynamic
tree models (DTC/DTL) were fit in a single (random) pass with $N=1000$
particles, and the number of MCMC iterations and restarts, etc., for
other Bayesian methods was such that Monte Carlo error in RMSE for
repeated runs on the same data is negligible against the results in
Figure \ref{f:fried}.  Non-Bayesian comparators were fit under package
defaults.  The dynamic versions of the tree models clearly outperform
their static counterparts, and the dynamic treed linear model (DTL)
performs nearly as well as or better than the GP and BART, both of
which offer flexible mean functions under a (true) constant variance
term.

\subsection{Search Optimization}
\label{sec:opt}

In \citet{TaddLeeGrayGrif2009},  TGP regression was used to
augment a local pattern search scheme with ranked lists of locations
that maximize a measure of expected improvement.  The GP leaf model
was successful in this setting -- deterministic
optimization with 8 inputs -- but pairwise covariance
calculations and repeated MCMC runs became unwieldy in higher
dimensions.  At the same time, such hybrid optimization schemes
are most useful in complicated high-dimensional settings:
although true global optimization is impossible without great expense,
a hybrid method uses regression to locate promising input areas and
move the local search appropriately, thus iterating towards robust
optimality. 

We propose that our dynamic regression trees, due to a flexible
on-line inference framework, provide a very attractive regression tool
for hybrid local-global search optimization. Although they are not
interpolators, and thus of limited usefulness in deterministic
optimization, both constant and linear regression trees provide an
efficient model for the prediction and optimization of stochastic
functions.  Such stochastic optimization problems are common in
operations research and, despite being deterministic, many engineering
codes are more properly treated as stochastic due to numerical
instability.  This section will outline use of dynamic regression
trees for the global component of a stochastic optimization search.

We seek to explore the input space in areas that are likely to provide
the optimum response (by default, a minimum).  In the deterministic
setting, \citet{JoneSchoWelc1998} search for inputs which maximize the
posterior expectation of an {\em improvement} statistic, $\mr{max}\{
f_{min} - f(\bm{x}), 0 \}$.  In more generality, improvement is just
the utility derived from a new observation.  For global search
optimization, there is no negative consequence of an evaluation that
does not yield a new optimum (this information is useful), and the
utility of a point which does provide a new optimum is, say, linear in
the difference between mean responses (other quantiles or moments are
also possible).  Thus, the improvement for a stochastic objective
$y(\bm{x})$ after $t$ observations is $I(\bm{x}) = \mr{max}\left\{
  \hat{y}_\mr{min} - \hat{y}(\bm{x}), 0 \right\}, $ where
$\hat{y}(\bm{x})$ is the regression model mean function (e.g.,
$\hat{y}(\bm{x}) = \mu_{\eta}$ for constant leaves or $\hat{y}(\bm{x})
= \mu_{\eta} + \bm{x}^\prime\beta_{\eta}$ for linear leaves) and
$\hat{y}_\mr{min}$ is this expected response minimized over the input
domain.

In hybrid schemes, we have found that maximizing $\ds{E}[I(\bm{x})]$
can be overly myopic.  We thus incorporate an active learning idea
(refer to Section \ref{sec:al}), and instead search for the maximizing
argument to $G(\bm{x}; \phi) = \ds{E}[I(\bm{x})] +
\mr{sd}(\hat{y}(\bm{x}))/\phi$, where $\phi$ is a precision parameter
that can be decreased to favor a more global scope.  Conveniently, use
of constant or linear leaves allows for both $\ds{E}[I(\bm{x})]$ and
$\mr{sd}(\hat{y}(\bm{x}))$ to be calculated in closed form conditional
on a given tree.  In particular, with $\mr{St}(a_\eta(\bm{x}),
b_\eta(\bm{x}), c_\eta )$ the posterior for $\hat{y}(\bm{x})$ given
tree $\mT_t$ and $\hat{y}_\mr{min}$ fixed at the minimum for
$a_\eta(\cdot)$ over our input domain, expected improvement
$\ds{E}[I]$ is (suppressing $\bm{x}$)
\[
(\hat{y}_\mr{min}-a_\eta)\mr{T}_{c_\eta}
\left( \frac{\hat{y}_\mr{min}-a_\eta}{\sqrt{b_\eta}}\right)
+\frac{\sqrt{b_\eta}}{c_\eta-1}\left[c_\eta + 
\frac{(\hat{y}_\mr{min}-a_\eta)^2}{b_\eta}\right]
\mr{t}_{c_\eta}\left( \frac{\hat{y}_\mr{min}-a_\eta}{\sqrt{b_\eta}}\right),
\]
with $\mr{T}_c$ and $\mr{t}_c$ the standard $t$ cumulative
distribution and density, respectively, with $c$ degrees of freedom
\citep[this is similar to the marginal improvement derived
in][]{WillSantNotz2000a}.  Finally, since $\ds{E}[I] = \frac{1}{N}
\sum_{i=1}^N \ds{E} [ I \mid\mc{T}^{(i)} ]$ and $\mr{\ds{V}
  ar}(\hat{y}) = \ds{E}[\mr{\ds{V}ar}(\hat{y} \mid \mc{T})] +
\mr{\ds{V}ar}(\ds{E}[\hat{y} \mid \mc{T}] )$, it is possible to obtain
$G =\ds{E}[I] + \mr{sd}(\hat{y})/\phi$ by evaluating the appropriate
functionals conditional on each $\mT \in \{\mT_t^{(i)}\}_{i=1}^N$
before taking moments across particles (see Section
\ref{sec:al} for further detail).

In a generic approach to sequential design, which is also adopted in
the active learning algorithms of Section \ref{sec:al}, the choice of
the ``next point'' for evaluation is based on criteria optimization
over a discrete set of candidate locations.  After initializing with a
small number of function evaluations, each optimization step augments
the existing sample $[\bm{x},y]^t$ by drawing a space-filling design
(we use uniform Latin hypercube samples) of candidate locations,
$\tilde{\bm{X}} = \{\tilde{\bm{x}}_i\}_{i=1}^M$, and finding
$\bm{x}^\star \in \tilde{\bm{X}}$ which maximizes $G(\bm{x}^\star)$.
Next, evaluate the new location, set $[\bm{x}_{t+1}, y_{t+1}] =
[\bm{x}^\star, y(\bm{x}^\star)]$, and follow the steps in Section
\ref{sec:pl} to obtain an updated particle approximation to the tree
posterior.  This is repeated as necessary.

\begin{figure}[p!]
\vskip -.5cm
\hskip -.5cm
\includegraphics[width=6.5in]{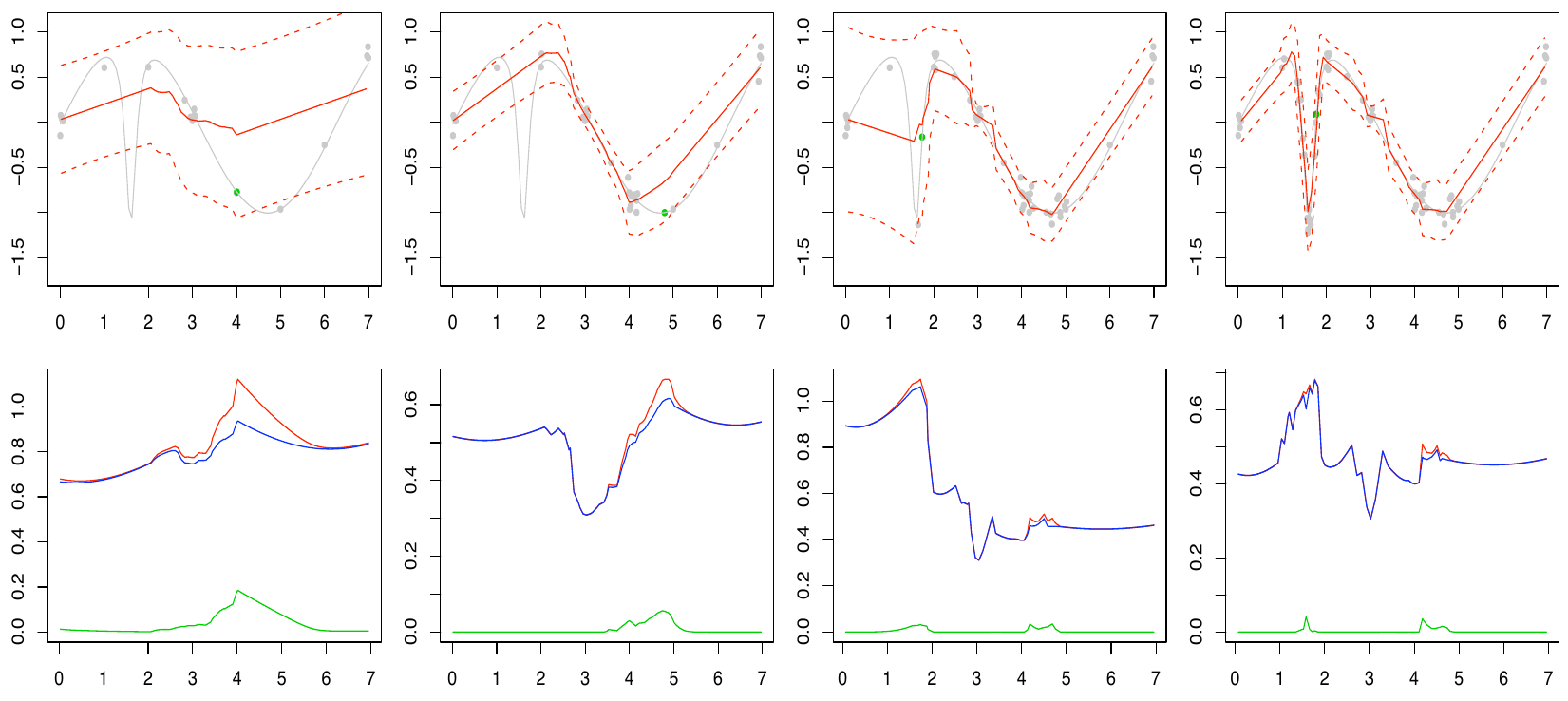}
\begin{center}
\vskip -.4cm
{\it \small Iterations 10, 20, 40, and 60 with $\phi = 1$.}
\end{center}
\hskip -.5cm
\includegraphics[width=6.5in]{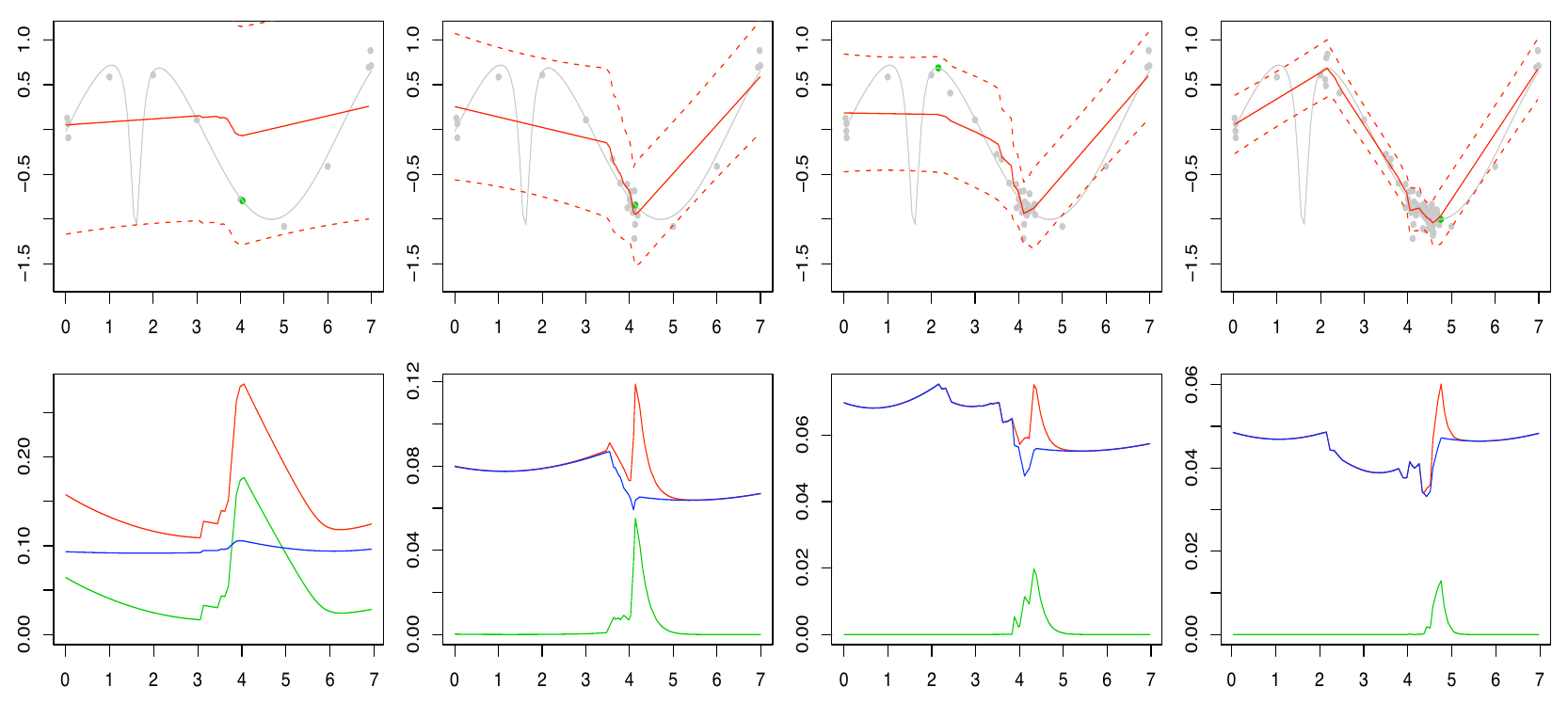}
\begin{center}
\vskip -.4cm
{\it \small Iterations 2, 15, 30, and 45 with $\phi = 10$.}
\end{center}
\caption{\label{fig:opt} Optimization with dynamic regression
  trees.  For each iteration, posterior mean and 90\% interval are on
  top (with true function and data in grey, and ``next point'' in
  green), while the bottom plot has $\ds{E}[I]$ in green,
  $\mr{sd}(\hat{y})$ in blue, and $G = \ds{E}[I]
  +\mr{sd}(\hat{y})/\phi$ in red.}
\end{figure}

Figure \ref{fig:opt} shows some results for mean optimization of the
test function $y = \mr{sin}(x) - \mr{Cauchy}(x; 1.6, 0.15) +
\varepsilon$, with $\varepsilon\sim \mr{N}(0,\sigma=0.1)$, for precision
values $\phi =1$ and $\phi =10$.  The fitted regression trees assume
linear leaves and, in each case, we used 500 particles and 100
candidate locations.  The role of $\phi$ in governing global scope is
clear, as the higher precision search is quickly able to locate a
robust minimum but does not then proceed with wider exploration.  The
linear leaf model seems to be efficient at fitting our test function.

\begin{table}[p!]
\vskip .5cm
\begin{center}
{\it  \hskip .5cm Constant  tree \hskip 2.75cm Linear tree \hskip 2.75cm Treed GP}\vskip .2cm
\begin{tabular}{|c|c|c||c|c|c||c|c|c|}\hline
$\phi=100$ &$\phi=10$ &$\phi=1 $&$\phi=100 $&$\phi=10$ &$\phi=1$ & $g=1$ & $g=5$& $g=10$\\
  \hline \hline
$-0.20$ & $-0.21$ & $-0.17$ & 
$-0.04$ & $-0.07$ & $-0.05$ & 
$-0.10$ & $-0.08$& $-0.13$ \\
$(0.13)$ & $(0.15)$ & 
$(0.15)$ & $(0.09)$ & $(0.13)$ & $(0.09)$ 
& $(0.12)$ & $(0.10)$ &  $(0.14)$\\
  \hline
\end{tabular}
\end{center}
\caption{\label{tab:opt} Minimization of the 2-d exponential function.  
  For each optimization scheme, the top number is mean solution
  and the bottom number is standard deviation (out of 50 runs).}
\end{table}

We also consider mean optimization of the two dimensional exponential
function $y \sim \mr{N}(x_1 \exp( -x_1^2 -x_2^2), 10^{-6})$, which is
discussed further in Section \ref{sec:al}.  Given data from a random
initial sample of 10 locations, our search routine was used to obtain
the next 10 function evaluations, with $\phi = 1$, $10$, or $100$,
using 1000 particles and candidate sets of 200 locations.  A TGP-based
search optimization, as described in \cite{GramTadd2010}, was also
applied.  This routine is the same as that outlined above for
dynamic trees, except that it maximizes $\ds{E}[I^g]$ rather than
$G$, where each iteration requires a new MCMC run (we use 500
iterations after a burn-in of 500), and the candidate set is augmented
with the location minimizing predicted response (such adaptation will
often lead to a more efficient search).  In this, the $g$ parameter
increases the global scope \citep[refer to][]{SchoWelcJone1998} and we
consider $g=1$, $5$, or $10$.  Each optimization scheme was repeated
50 times, and the summary is shown in Table \ref{tab:opt}.

In a result that will parallel findings in Section \ref{sec:al},
constant, rather than linear, leaves lead to a better exploration of
the 2-d exponential function; this relative weakness exposes a
difficulty for linear models in fitting the rapid oscillations around
$(0,0)$.  Interestingly, constant trees also lead to better results
than for the TGP optimization routine.  This is despite the added
flexibility of GP leaves, an augmented candidate sample, and use of a
wide range of $g$ values for $\ds{E}[I^g]$ (which, although harder to
compute when marginalizing over model parameters, is the literature
standard for this type of search).  In addition, due to repeated MCMC
runs, the TGP algorithm requires near to 10 times the computation of
sequential tree optimization.

\subsection{Active Learning}
\label{sec:al}

Another common application of sequential design is focused on
evaluating an unknown response surface -- the mean function
$f(\bm{x})$ in (1.a) -- at its most informative input location.  Such
{\it active learning} procedures are intended to provide efficient
automatic exploration of the covariate space, thus guiding an on-line
minimization of prediction error.  As in the search optimization of
Section \ref{sec:opt}, each active learning iteration will draw
candidate locations, $\tilde{\bm{X}} = \{\tilde{\bm{x}}_i\}_{i=1}^M$,
and select the next ($t+1^{\mathrm{st}}$) design point to be $\bm{x}^\star \in
\tilde{\bm{X}}$ which maximizes a heuristic statistic.  However,
instead of aiming for a function optimum, we are now solely interested
in understanding the response surface over some specified input range.
Hence, we need to choose $\bm{x}^\star$ to maximize some measure of
the predictive information gained by sampling
$y(\bm{x^\star})$.

Two common heuristics for this purpose are active learning
\citet[ALM;][]{mackay:1992} and active learning
\citet[ALC;][]{cohn:1996}.  An ALM scheme selects the $\bm{x}^\star$
that leads to maximum variance for $y(\bm{x}^\star)$, whereas ALC
chooses $\bm{x}^\star$ to maximize the expected reduction in
predictive variance averaged over the input space.  A comparison
between approaches depends upon the regression model and the
application, however it may be shown that both approximate a maximum
expected information design and that ALC improves upon ALM under
heteroskedastic noise. Both heuristics have computational demands that
grow with $|\tilde{\bm{X}}|$: ALC requires time in
$O(|\tilde{\bm{X}}|^2)$, wheres ALM is in $O(|\tilde{\bm{X}}|)$.
\citet{gra:lee:2009} provide further discussion of active learning as
well as extensive results for ALC and ALM schemes based on MCMC
inference with CGM's static treed constant/linear (TC/L) models, GPs,
and TGPs.

Since active learning is an inherently on-line algorithm, it provides
a natural application for dynamic regression trees.  The remainder
of this section will illustrate ALC/ALM schemes built around particle
learning for trees, and show that our methods compare favorably to
existing MCMC-based alternatives.  As in \ref{sec:opt}, both constant
and linear leaf models lead to closed-form calculations of heuristic
functionals conditional on a given tree.  Fast prediction allows us to
evaluate the necessary statistics across candidates, for each particle,
and hence  find the optimal $\bm{x}^\star$ under  our
heuristic.  Trees are then updated for $y(\bm{x}^\star)$, and the
process is repeated.

ALM seeks to maximize $\mr{\ds{V}ar}(y(\bm{x}))$, which is fairly
simple to calculate. For a given tree $\mT_t$ such that $\bm{x}$ is
allocated to leaf node $\eta \in L_{\mT_t}$, let
$\mu_{\eta}(\bm{x})=\ds{E}[y(\bm{x})\mid \eta]$ and $\mr{v}_{\eta}(\bm{x})
= \mathrm{\ds{V}ar}(y(\bm{x}) \mid \eta) $ denote the conditional
predictive mean and variance respectively for $y(\bm{x})$.  These are available
from equations (\ref{eq:cpred}) and (\ref{eq:lpred}) for each leaf
model as
$\mu_{\eta}(\bm{x}) = \bar{y}_{\eta}$ or $\mu_{\eta}(\bm{x}) = \bar{y}_{\eta} +
\hat{\bm{x}}^\prime\hat{\beta}_{\eta}$ and
$
\mr{v}_{\eta}(\bm{x}) =  s_{\eta}^2 (1 + {1}/{|\eta|})/ (|\eta| - 3)$ or
$\mr{v}_{\eta}(\bm{x}) =[s_{\eta}^2-\mathcal{R}_{\eta}]\left( 1 + {1}/{|\eta|} + 
        \hat{\bm{x}}^\prime 
        \mathcal{G}_{\eta}^{-1} \hat{\bm{x}}
      \right)/( |\eta| - d - 3 )$.
 Given the  particle set $\{\mT_t^{(i)}\}_{i=1}^N$,
the unconditional predictive variance is
\vspace{-.4cm}
\begin{align}\label{varid} 
\mr{\ds{V}ar}(y(\bm{x})) &= 
  \ds{E}[\mathrm{\ds{V}ar}(y(\bm{x})\mid \mT)] +
  \mathrm{\ds{V}ar}(\ds{E}[y(\bm{x})\mid \mT]) \nonumber \\
  &\approx \frac{1}{N} \left[\sum_{i=2}^N \mr{v}^{(i)}_{\eta}(\bm{x}) +
    \mu^{(i)}_\eta(\bm{x})^2 \right] -
  \left[\frac{1}{N}\sum_{i=1}^N \mu^{(i)}_\eta(\bm{x})\right]^2,
\end{align}
which is straightforward to evaluate for all $\bm{x} \in
\tilde{\bm{X}}$ during our search for the maximizing $\bm{x}^\star$.

The ALC statistic is more complicated.  Let $\Delta
\sigma^2_{\bm{x}}(\bm{x}') = \mr{\ds{V}ar}(y(\bm{x}')) -
\mr{\ds{V}ar}(y(\bm{x}') \mid \bm{x}')$ denote the reduction
in variance at $\bm{x}'$ when the design is augmented with $\bm{x}$.
Since $y(\bm{x})$ is not observed, the ALC statistics must condition
on the existing model state.  Hence, we define $\Delta
\sigma^2_{\bm{x}}(\bm{x}') = \ds{E}[ \Delta \sigma^2_{\bm{x}}(\bm{x}'
\mid \mT_t)] $ at time $t$ in terms of the conditional variance
reduction
\begin{equation}
  \Delta \sigma^2_{\bm{x}}(\bm{x}' \mid \mT_t ) = 
  \mr{\ds{V}ar}(y(\bm{x}')\mid\mT_t) - 
  \mr{\ds{V}ar}(y(\bm{x}')\mid \bm{x}, \mT_t).
\end{equation}
Expressions for $\Delta \sigma^2_{\bm{x}}(\bm{x}' \mid 
\mT_t)$ may be obtained as a special case of the results given by
\cite{gra:lee:2009}.  Clearly, $\Delta \sigma^2_{\bm{x}}(\bm{x}'\mid
\mT_t) = 0$ if $\bm{x}$ and $\bm{x}'$ are allocated to different
leaves of $\mT_t$.  Otherwise, for $\bm{x}$ and $\bm{x}'$
allocated to $\eta \in L_{\mT_t}$ in a given tree, we have that
$\Delta \sigma^2_\bm{x}(\bm{x}'\mid \mT_t) = \Delta
\sigma^2_{\bm{x}}(\bm{x}'\mid\eta)$ is (for constant and linear
leaves respectively)
\begin{eqnarray}
\frac{s_{\eta}^2}{|\eta|-3} \times
\frac{\left(\frac{1}{|\eta|}\right)^2}{1 + \frac{1}{|\eta|}}
~~~~~\mbox{or} ~~~~~
\frac{s_{\eta}^2 - \mathcal{R}_{\eta}}{|\eta|-m-3} \times
\frac{\left(\frac{1}{|\eta|} +(\hat{\bm{x}}')^\prime \mathcal{G}^{-1}_{\eta}\hat{\bm{x}}\right)^2}{
1 + \frac{1}{|\eta|} +(\hat{\bm{x}}')^\prime \mathcal{G}^{-1}_{\eta}\hat{\bm{x}}}, \label{eq:lmalc}
\end{eqnarray}
where $\hat{\bm{x}} = \bm{x}-\bar{\bm{x}}_\eta$ and 
similarly for $\hat{\bm{x}}'$.  In practice, the integral 
$\Delta \sigma^2(\bm{x}) = \int_{\mathbb{R}^d} \Delta \sigma^2_{\bm{x}}(\bm{x}')\,d\bm{x}'$ 
is approximated by a discrete sum over the random space-filling set 
$\tilde{\bm{X}}$.  Hence, $\bm{x}^\star$ is the candidate location 
which maximizes the sum of $\Delta \sigma^2_{\bm{x}^\star}(\bm{x}'\mid \mT_t) $ 
over both $\bm{x}' \in \tilde{\bm{X}}$ and $\mT_t \in \{\mT_t^{(i)}\}_{i=1}^N$.

\begin{figure}[t]
\centering
\includegraphics[scale=0.6,trim=0 73 0 0,clip=TRUE]{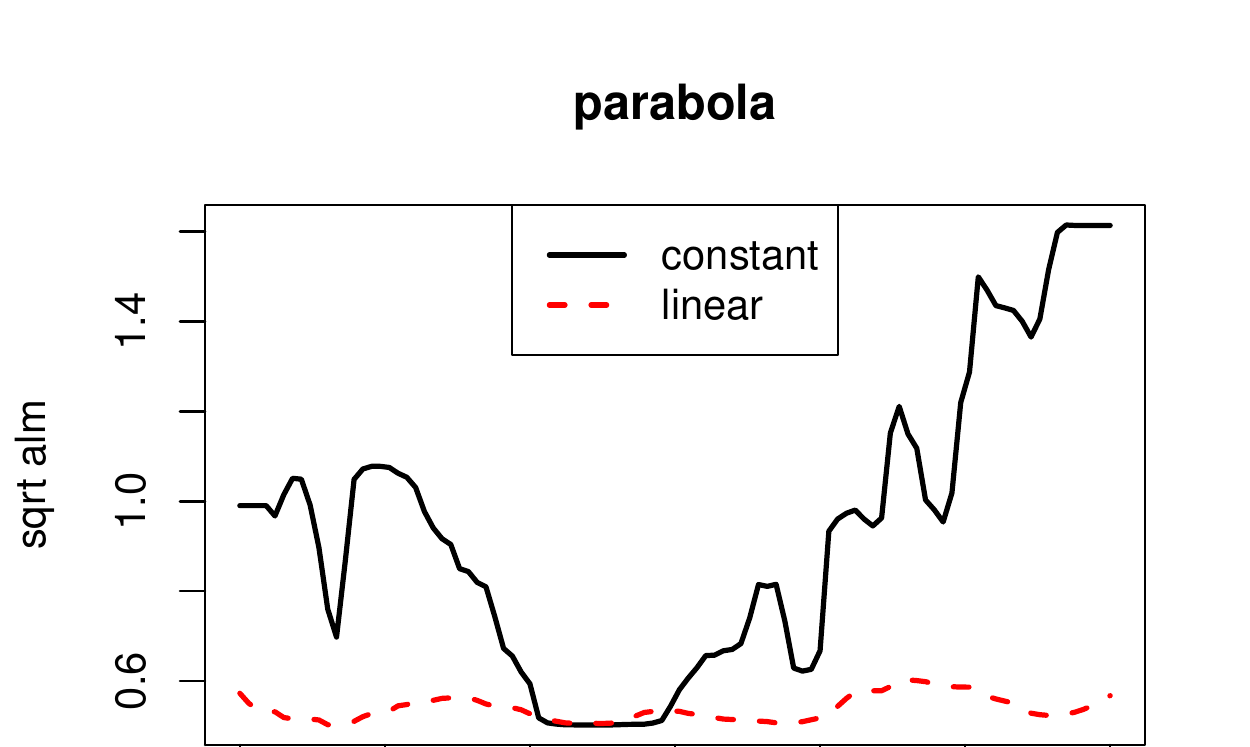}
\includegraphics[scale=0.6,trim=25 73 0 0,clip=TRUE]{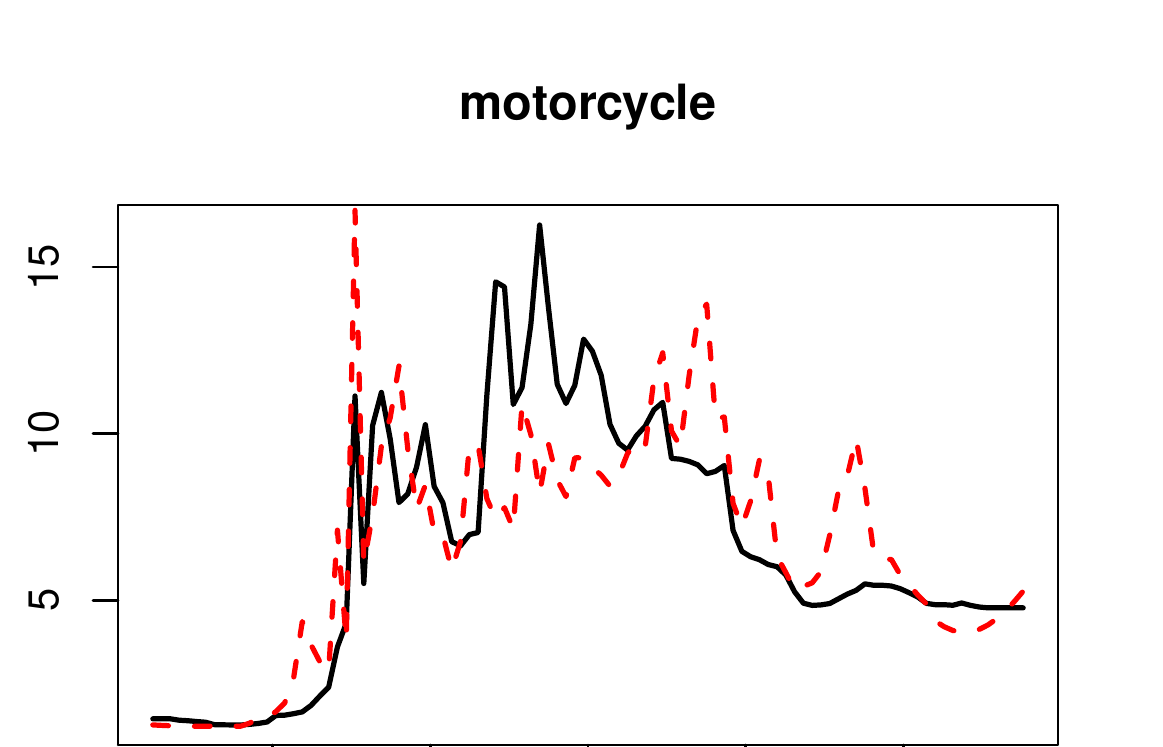}
\includegraphics[scale=0.6,trim=0 10 0 40,clip=TRUE]{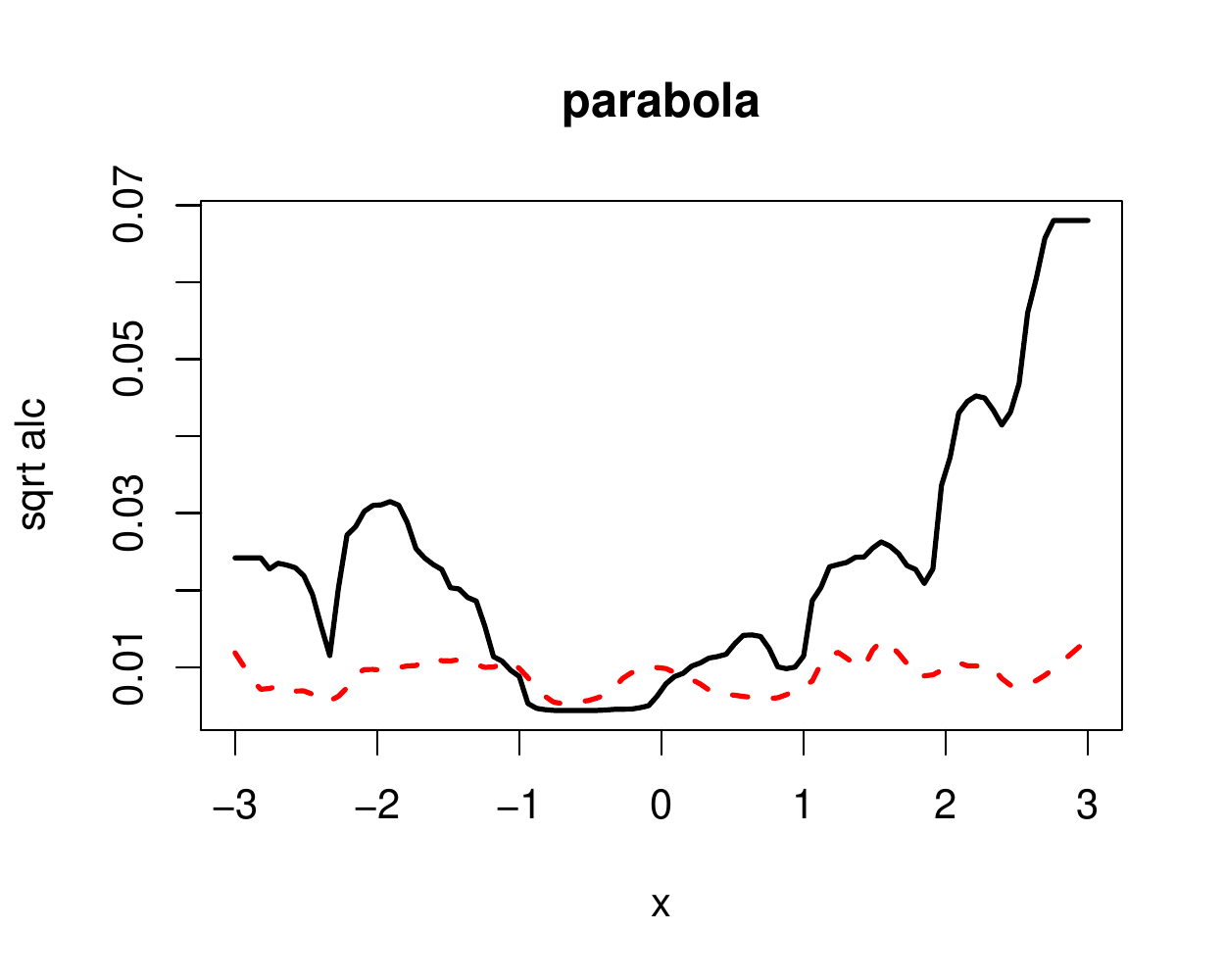}
\includegraphics[scale=0.6,trim=25 10 0 40,clip=TRUE]{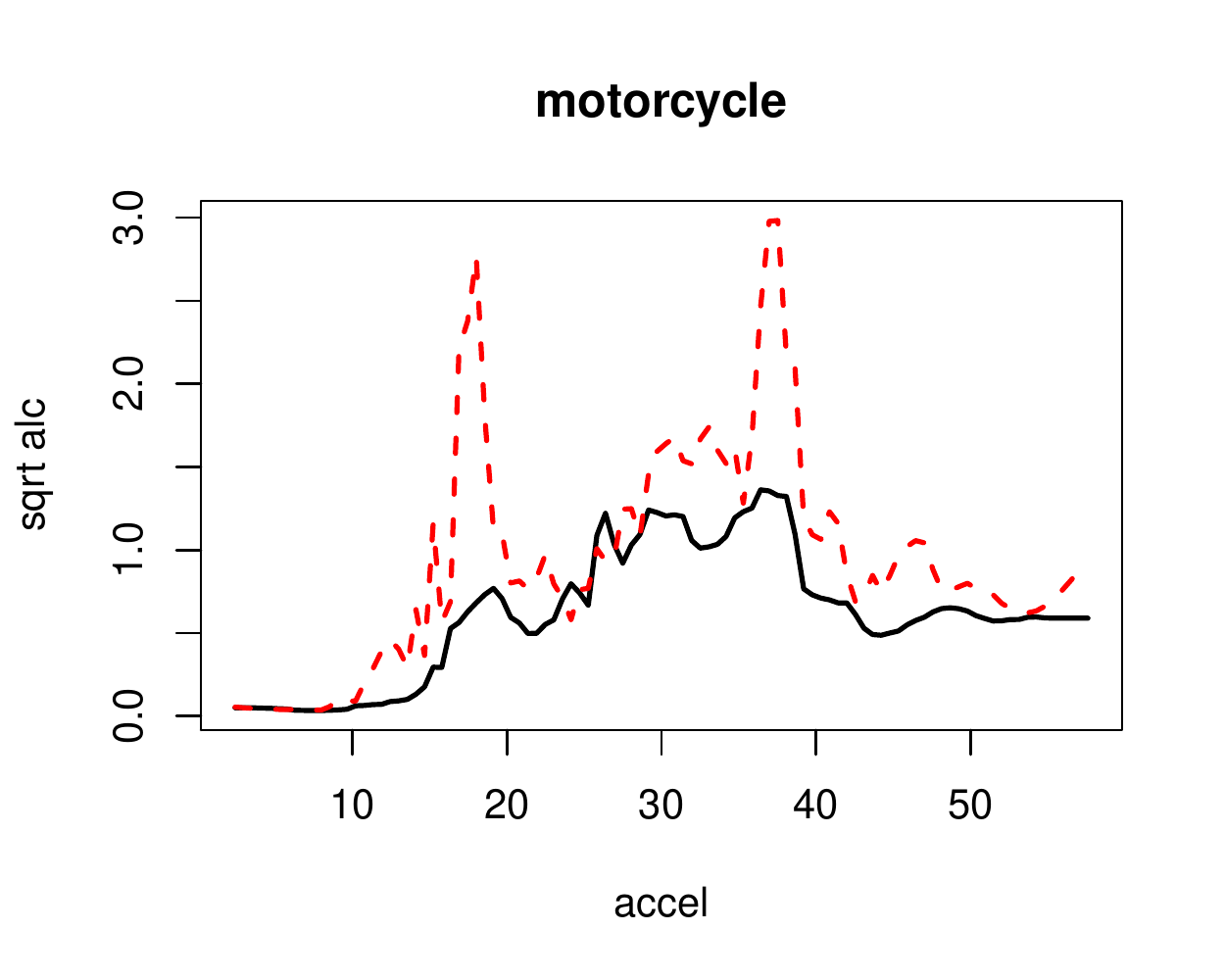}
\caption{Active learning comparison for the parabola (left) and
  motorcycle data (right).  The top row shows $\mr{\ds{V}ar}(y(\bm{x}))$ for ALM
  and the bottom shows $\sqrt{\Delta \sigma^2(\bm{x})}$ for ALC.}
\label{f:actlearn}
\end{figure}

We use the parabola and motorcycle data examples of Section
\ref{sec:reg} to compare ALC and ALM for both constant and linear leaf
dynamic trees.  Figure \ref{f:actlearn} illustrates the statistics associated
with each combination of the two heuristics and two regression 
models, evaluated given the complete datasets.
For the parabola data (left column), the ALM and ALC plots look roughly similar 
and, in each case, the linear model statistics are much more 
flat than for the constant model.  In contrast, the ALM and ALC plots are 
very different from each other for the  motorcycle data, 
which exhibit heteroskedastic additive error.

\begin{figure}[p]
\centering
\vspace{0.25cm}
\begin{tabular}{ccc}
\ \ \ \ $t=10$ & \ \ \ \ $t=25$ & \ \ \ \ $t=55$ \\
\includegraphics[scale=0.5,trim=30 0 0 0]{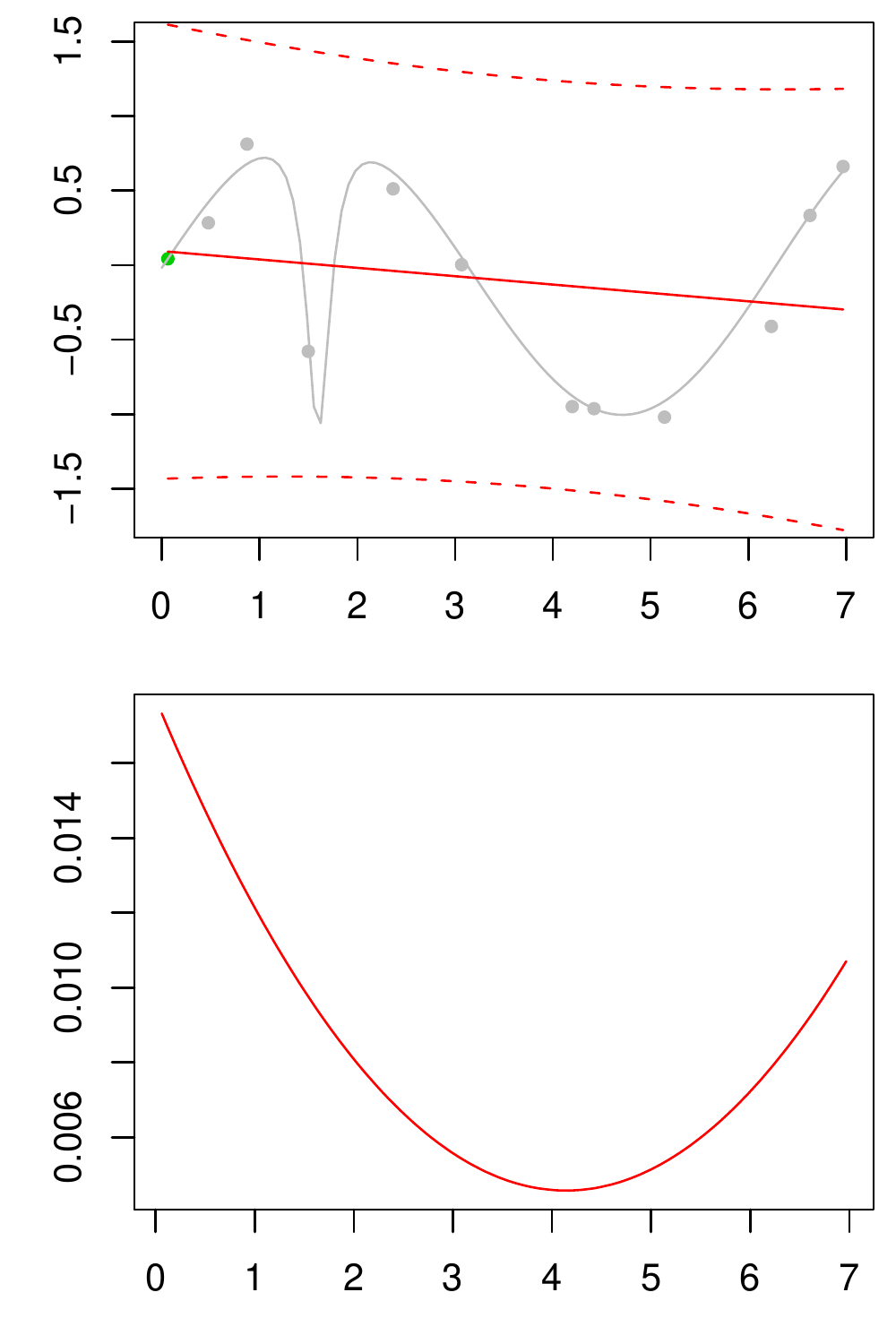} &
\includegraphics[scale=0.5,trim=25 0 0 0]{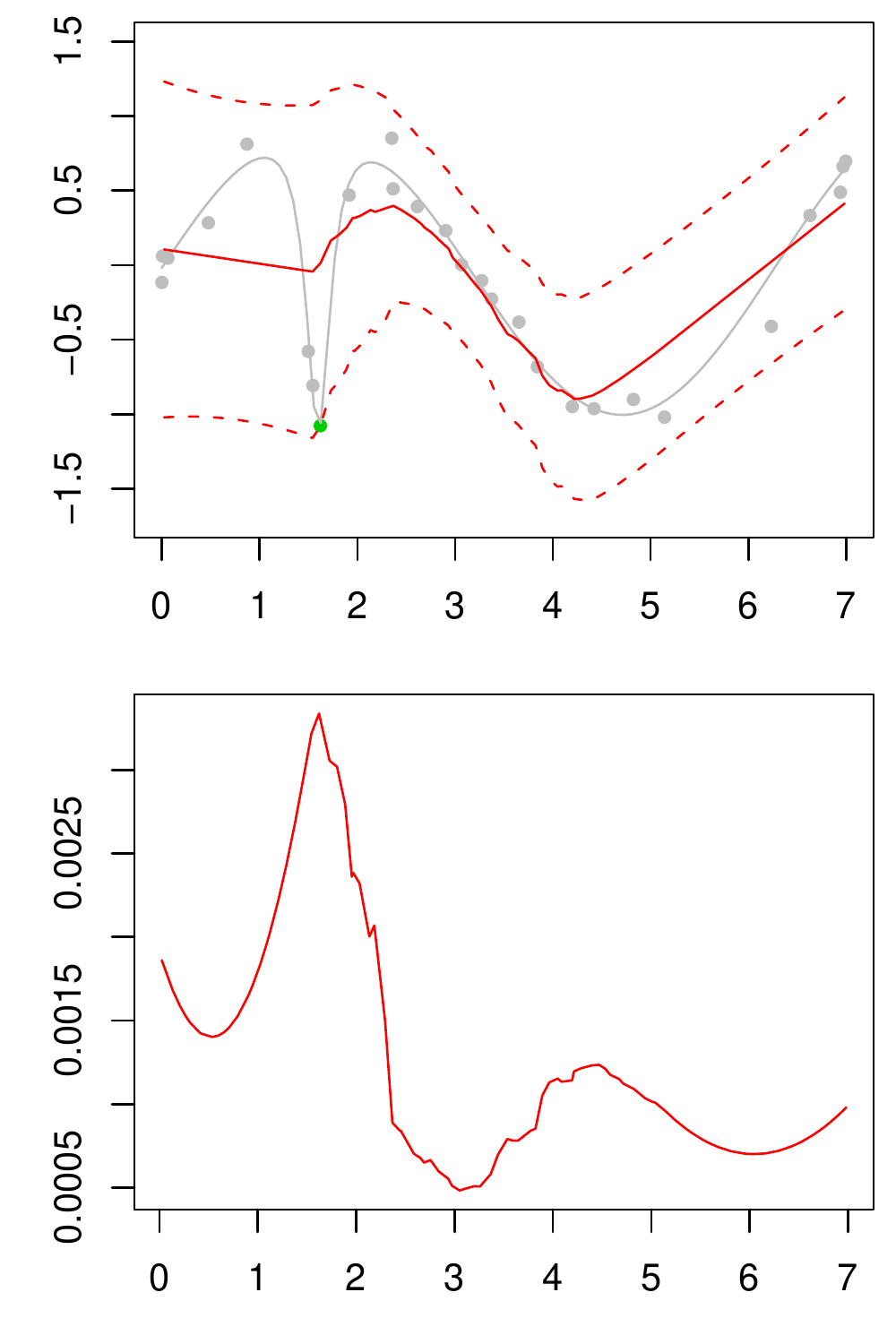} &
\includegraphics[scale=0.5,trim=25 0 0 0]{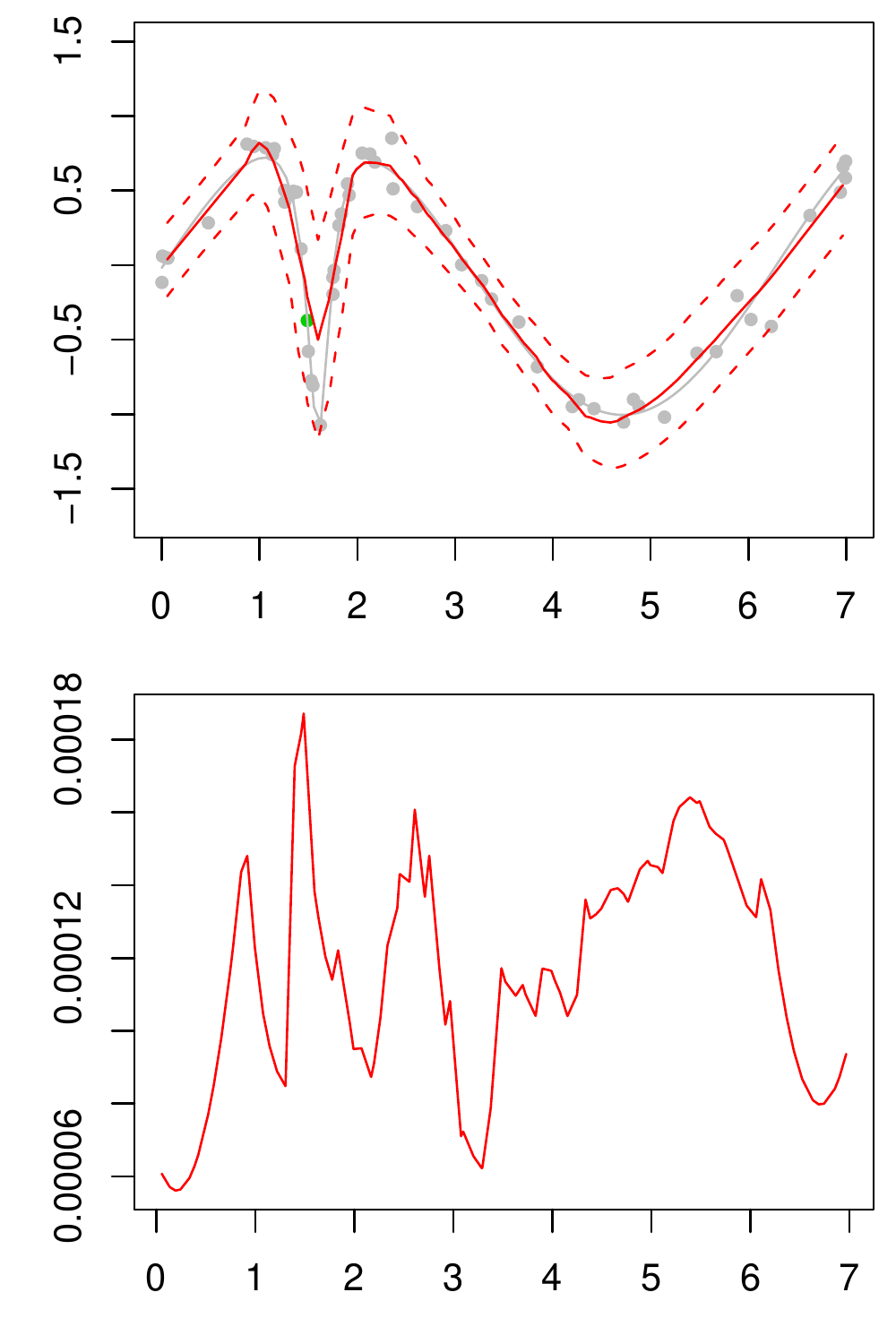}
\end{tabular}
\caption{Snapshots showing the progression of the ALC heuristic driven
  by the dynamic tree model with linear leaves on the sin/Cauchy
  test data.  The top panels show the true surface (grey lines),
  sampled input/output pairs (grey points), and the dynamic tree
  predictive means and 90\% quantiles (red lines).  The bottom panels
  show the corresponding $\Delta \sigma^2(\bm{x})$ surface.}
\label{f:pltlalc}
\end{figure}

\begin{table}[p]
\begin{center}
\begin{tabular}{rcr}
\multicolumn{3}{c}{\bf 1-d Sin/Cauchy} \\
  \hline
  model  & design & RMSE \\
  \hline
 TGP & ALC & 0.0641 \\ 
 TGP & ALM & 0.0660\\
 DTL & ALC & 0.0755\\
 GP & ALM & 0.0847\\
 DTL & ALM & 0.0992\\
 -- &  LHS & 0.1251\\
 GP & ALC & 0.1265\\
 -- &  ME & 0.1325\\
 TC & ALC & 0.1597\\
 DTC & ALC & 0.1616\\
 TC & ALM & 0.1960\\
 TL & ALC & 0.2193\\
 DTC & ALM & 0.2251\\
 TL & ALM & 0.2458\\
   \hline
\end{tabular}
\hspace{2cm}
\begin{tabular}{rcr}
\multicolumn{3}{c}{\bf 2-d Exponential} \\
  \hline
model & design & RMSE \\
  \hline
TGP & ALM & 0.00577 \\ 
DTC & ALC & 0.00732\\
DTL & ALM & 0.00740\\
TGP & ALC & 0.00852\\
DTL & ALC & 0.00936\\
DTC & ALM & 0.00997\\
TC & ALM & 0.01126\\
TC & ALC & 0.01182\\
TL & ALM & 0.01622\\
GP & ALM & 0.02411\\
TL & ALC & 0.02431\\
-- & LHS & 0.03636\\
-- & ME & 0.04408\\
GP & ALC & 0.05419\\
   \hline
\end{tabular}
\end{center}
\caption{Comparing models and active learning heuristics on
  the 1-d sin/Cauchy data and the 2-d exponential
  data.  The tables are sorted on the fourth column (RMSE).  }
\label{t:alcompare}
\end{table}

For a further comparison we return to the sin/Cauchy data from Section
\ref{sec:opt}.  Figure \ref{f:pltlalc} shows 
sequential design progress under our dynamic tree model with
linear leaves and the ALC heuristic in three snapshots: after an
initial $t=10$ Latin hypercube (LHS) sample, after 15 samples taken via
ALC ($t=25$) and after 45 ALC samples ($t=55$).  With the exception of
$t=10$, when there is not enough data to support a split in the
tree(s), the ALC statistic is high at points where $y(\bm{x})$ is changing
direction---highest where it is changing most rapidly.

Table \ref{t:alcompare} (left) shows how the dynamic tree models
compare to various static Bayesian treed models, including TGP, on
this data.  All experiments were started with an initial LHS in
$[0,7]$ of size 10 followed by 40 active learning rounds.  Each round
uses 20 random LHS candidates $\tilde{\bm{X}}$ in $[0,7]$. At the end
of the 40 rounds the root mean squared error (RMSE) was calculated for
a comparison of predictive means to the truth in a size 200 hold-out
set from a GP-based maximum entropy (ME) design in $[0,7]$.  This was
repeated 30 times and the average RMSE is shown in the table.  The
expensive TGP methods dominate, with dynamic trees using linear leaves
(DTL) trailing.  In both cases, ALC edges out ALM.  Interestingly,
while static treed linear models (TL) fit through MCMC are amongst the
worse performers, their sequential alternatives are amongst the best.
Neither TC or DTC are strong performers, perhaps due to the almost
linear derivatives of our test function.  Since the ``interesting''
aspects of this response are evenly distributed throughout the input
space, both the GP with ALM and the offline LHS and ME comparators
also do well.

Table \ref{t:alcompare} (right) shows the results of a similar
experiment for the 2-d exponential data from Section \ref{sec:opt}.
The setup is as described above, except that the initial LHS is now of
size 20 and is followed by 55 active learning rounds.  Here, the
dynamic trees share the top 6 spots with TGP only, and clearly
dominate their static analogues. Indeed, DTC is the second best
performer, echoing our findings in Section \ref{sec:opt} that the
constant leaf model is a good fit for this function.  In contrast to
the results from our simple 1-d example, the offline (LHS and ME) and
GP methods are poor here because the region of interest is confined to
the bottom left quadrant of our 2-d input space.  Finally, in all of
these examples, dynamic regression trees (DTC and DTL) are the
only methods that run on-line and do not require batch MCMC
processing.

\subsection{Classification}
\label{sec:class}

Classification is one of the original applications for tree models,
and is very often associated with sequential inference
settings. Categorical data can be fit with multinomial leaf trees
(as in \ref{sec:mnl}) and, without an application specific loss
function, the predictive classification rule assigns to new inputs the
class with highest mean conditional posterior probability.  That is,
\begin{equation}\label{eq:classrule}
\hat{c}(\bm{x}) = \mr{arg~max}\left\{\frac{1}{N}
\sum_{i=1}^N\left( \hat{p}^{\eta(\bm{x})}_c\right)^{(i)}:~{c=1,\ldots,C} \right\},
\end{equation}
for a set of $N$ particles, where each $\hat{p}^{\eta(\bm{x})}_c$ is
the $c$-class probability from (\ref{eq:clpred}) for the leaf
corresponding to $\bm{x}$.  Evaluating $\hat{c}(\bm{x})$ over the
input space provides a predictive classification surface.

We first apply our methodology to the Cushing's data, again available
in the {\tt MASS} library for {\sf R}, used throughout the book by
\citet{Ripl1996} to illustrate various classification algorithms.
This simple data set has two inputs -- each patient's urinary
excretion rates for tetrahydrocortisone and pregnanetriol, considered
on the log scale -- and the response is one of three different types of
Cushing's syndrome (referred to as $a$, $b$, or $c$).  We fit the
multinomial leaf dynamic tree model to this data, and compare our
results to those from a GP soft-max classifier as outlined in
\cite{BrodGram2009}, which fits independent GP models to each latent
response $y_c$ such that $p_c = \mr{exp}({y_c})/\sum_{l=1}^C
\mr{exp}({y_l})$.

\begin{figure}[t]
\vskip -.5cm
\includegraphics[width=6.3in]{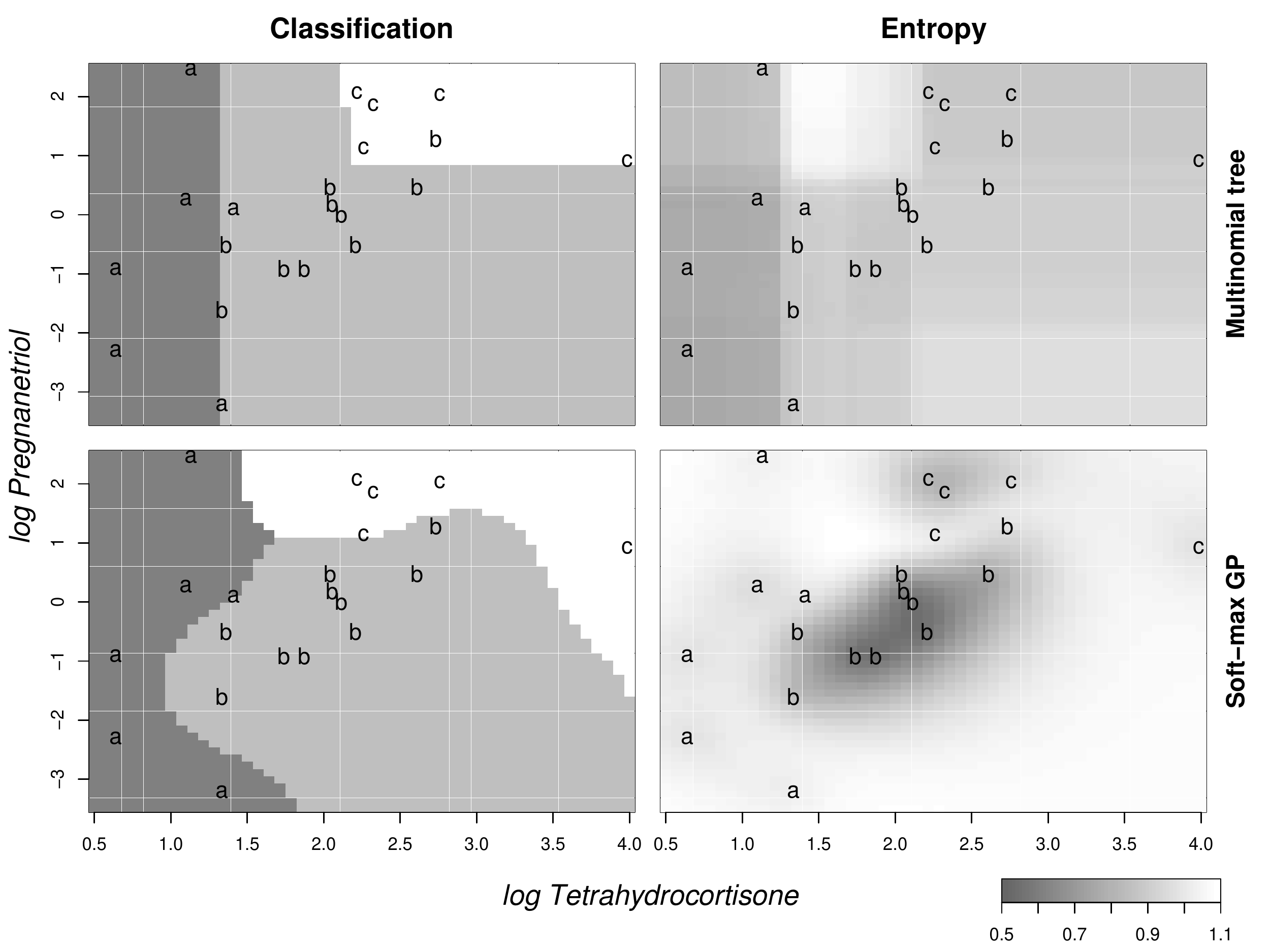}
\vskip -.3cm
\caption{\label{fig:class} Cushing's data example.  The left column shows
  the class corresponding to maximum mean posterior probability, and
  the right has entropy for the posterior mean probability vector.
  Results for a dynamic classification tree are on top, and for a
  GP soft-max classifier on the bottom.  Class (a$\rightarrow$c) and
  entropy rise from black to white, and plotted data is labelled by
  class. }
\end{figure}

Figure \ref{fig:class} illustrates the outcome of this example.  The
dynamic tree results (top row) are based on a set of 1000
particles, while the GP soft-max results (bottom row) use every
100$^\mr{th}$ observation of a 10,000 iteration MCMC chain.  The left
column of this figure shows the syndrome classification surface based
on maximum mean posterior probability, as in (\ref{eq:classrule}).
Although it is not possible to assess whether any classifier is
outperforming the other, these surfaces clearly illustrate the effect
of axis-symmetric (tree) classification as opposed to the
surface from a radial-basis (GP) model.  We also note that the GP
soft-max classification rule is very similar to results from
\citet[][chap. 5.2]{Ripl1996} for a neural network fit with weight
decay.

The right column of Figure \ref{fig:class} shows the entropy,
$-\sum_{l \in \{a,b,c\}} p_l \log(p_l)$, of the posterior mean
probability surface for each classification model.  Such entropy
plots, which are commonly used in classification settings to assess
response variability, illustrate change in expected value of
information over the input space.  Indeed, \citet{GramPols2009}
propose entropy as an active learning criteria for classification
problems, analogous to ALC in Section \ref{sec:al}.  In our
application, dynamic tree entropy is peaked around a 3-class
border region in the top-left.  This intuitively offers better
guidance about the value of additional information than the GP entropy
surface, which is high everywhere except right at clusters of
same-class observations.

To better assess the performance of our dynamic tree classifier, we
turn to a larger ``credit approval'' data set from the UC Irvine
Machine Learning Database.  This data includes 690 credit card
applicants grouped by approval status (either `+' or `-') and 15 input
variables. Eleven of these inputs are categorical, and for each of
these we have encoded the categories through a series of binary
variables.  This convenient reparametrization, which expands the input
space to 47 dimensions, allows direct application of the model in
Section \ref{sec:trees} with nodes splitting on $x_i=0$ or $1$ for
each binary variable \citep[see][for discussion of this approach in
general partition trees]{GramTadd2010}.  As an aside, we note that the
binary encoding can also be used to incorporate categorical data into
constant and linear leaf trees; the only adaptation is, for linear
leaves, to exclude binary variables from each regression design
matrix.

We applied the multinomial dynamic tree to 100 independent repetitions
of training on 90\% of the credit approval data and predictive
classification on the left-out 10\% sample.  This experiment is
identical to that in \cite{BrodGram2009}, who tested both a TGP
soft-max classifier (includes binary variables for partitioning of
latent TGP processes and fits independent GP to continuous inputs
within each partition) and a na{\"i}ve GP soft-max classifier (fit to
the reparametrized 47 inputs, treating binary variables as continuous
in the correlation function).  Their results are repeated here, beside
those for our dynamic tree, in Table \ref{tab:class}.  Once again,
despite using a less sophisticated leaf model, the dynamic tree is a
clear performance winner and the only classifier able to beat 14\%
average missclassification.  The dynamic tree was also
orders-of-magnitude faster than the GP and TGP classifiers (we use
1000 particles, and they have 15,000 iterations), and only our
approach will be feasible in on-line applications.

\begin{table}[h!]
\vskip .5cm
\begin{center}
\begin{tabular}{|c||c|c|c|}\hline
{\it Classifier} & {\it Dynamic Multinomial Tree} & 
{\it TGP soft-max} & {\it GP soft-max}\\
  \hline 
Missclassification rate & 0.136 (0.038) & 0.142 (0.036) & 0.146 (0.04)\\
Time (CPU hours) per fold & 0.01 & 1.62 & 5.52\\
  \hline
\end{tabular}
\end{center}
\vskip -.5cm
\caption{\label{tab:class} 
  Mean (and standard deviation) for out-of-sample 
  missclassification over 10 random 10-fold cross-validations 
  and computing time for each classifier on the credit approval
  data. }
\vskip -.2cm
\end{table}

\section{Discussion}
\label{sec:discuss}

We have reformulated regression trees as a dynamic model.  This
sequential characterization leads to an entirely new class of models
which can be fit on-line; have a scale-free automatic prior
specification; avoid the need for parameter sampling; and lead to
robust and flexible prediction. We have shown empirically that our
dynamic regression trees can provide superior performance and are
less expensive than common alternatives.

In a key point, we have been able to define particles for filtering
which contain only split rules and sufficient statistics, leading our
sequential filtering to efficiently discard all partition models
except those which are predicting well.  Due to the size and
complexity of potential tree space, this narrowing of posterior search
is an essential aspect of our models' success. In addition,
restrictions on partition size allow us to make use of improper priors
(for constant and linear leaves), which is not usually possible in
particle inference.

The modeling and examples contained herein provide encouragement for
further work under a strategy that looks to take advantage of 
sequential model characterizations.  In particular, we
hope to find that similar ideas on update mechanisms
for covariate dependent model features will lead to insight about
dynamic versions of other graphical structures.

\hlf \bibliography{plt}
\bibliographystyle{jasa}

\end{document}